\documentclass[10pt,journal,twocolumn]{IEEEtran}
\usepackage[T1]{fontenc}

\usepackage[applemac]{inputenc}

\ifCLASSINFOpdf
  
\else
 
\fi

\usepackage{epsfig}
\usepackage{amsmath}
\usepackage{amsthm}
\usepackage{amsfonts}
\usepackage{amssymb}
\usepackage{hyperref}
\hypersetup{colorlinks,%
citecolor=black,%
filecolor=black,%
linkcolor=black,%
urlcolor=black,%
pdftex} 
\usepackage{graphicx}
\usepackage{graphics}
\usepackage{float}
\usepackage{url}
\usepackage{setspace}
\usepackage{subfigure}
\usepackage{balance}
\usepackage{amsmath} 
\usepackage{bm}      

\usepackage{color}
\usepackage{algpseudocode}
\usepackage{algorithm}
\usepackage{stfloats}
\usepackage{lipsum}
\usepackage{float}
\usepackage[normalem]{ulem}
\useunder{\uline}{\ulined}{}%
\hypersetup{
    colorlinks=blue,
    linkcolor=blue,
    filecolor=blue,      
    urlcolor=blue,
}

\def\mi0{\mathbf{0}}
\def\mih{\mathbf{h}}

\def\mix{\mathbf{x}}

\def\ontop#1#2{\setbox0\hbox{#2}\copy0\llap{\raise\ht0\hbox{#1}}}

\usepackage{etoolbox}
\AtBeginEnvironment{matrix}{\setlength{\arraycolsep}{1.2pt}} 

\makeatletter
\renewcommand*\env@matrix[1][\arraystretch]{%
  \edef\arraystretch{#1}%
  \hskip -\arraycolsep
  \let\@ifnextchar\new@ifnextchar
  \array{*\c@MaxMatrixCols c}}
\makeatother

\usepackage{hyperref}
\begin{document}

\title{Dynamic Power Allocation For NOMA-Based Transmission in 6G Optical Wireless Networks}

\author{\IEEEauthorblockN{Ahmad Adnan Qidan, Taisir El-Gorashi, Majid Safari, Senior
Member, IEEE, Harald Haas, Fellow, IEEE,  Richard V. Penty, Fellow, IEEE,  Ian H. White, Fellow, IEEE, and Jaafar M. H. Elmirghani, Fellow, IEEE}

\thanks{ Ahmad Adnan Qidan, Taisir E. H. El-Gorashi and Jaafar M. H. Elmirghani are with Department of Engineering, Faculty of Natural, Mathematical and Engineering Sciences, King's College London, London, United Kingdom (UK). (e-mails: ahmad.qidan@kcl.ac.uk; taisir.elgorashi@kcl.ac.uk; jaafar.elmirghani@kcl.ac.uk)

Majid Safari is with  the School of Engineering, Institute for Digital Communications, The University of Edinburgh, Edinburgh, UK. (e-mail: majid.safari@ed.ac.uk).

Harald Haas is with the LiFi Research and Development Centre (LRDC), Electrical Engineering Division, Department of Engineering, University of Cambridge, Cambridge, UK. (e-mail: huh21@cam.ac.uk).

Richard V. Penty is with the Centre for Photonic Systems, Electrical Engineering Division, Department of Engineering, University of Cambridge, Cambridge, UK. (e-mail: rvp11@cam.ac.uk).

Ian H. White is with the University of Bath, Bath, UK. (e-mail:
i.h.white@bath.ac.uk).}
}

\markboth{SUBMITTED TO IEEE  XXX~2025}
{Khulood \MakeLowercase{\emph{et al.}}: hh}

\def\be{\begin{equation}}
\def\ee{\end{equation}}

\maketitle

\begin{abstract}
Optical wireless communication (OWC) has been considered as a key enabling technology to unlock unprecedented speeds of communication, supporting high demands of data traffic. In this paper, infrared lasers are used as optical transmitters operating in an indoor environment under eye safety regulations  due to their high modulation speed. To provide efficient multiple access service, non-orthogonal multiple access (NOMA)-based transmission is implemented to multiplex  messages intended to multiple users in the power domain and maximize the spectral efficiency of our laser-based OWC network. In particular, a blind interference alignment (BIA) outer precoder is designed to coordinate the transmission among multiple optical access points (APs) and determine the precoding matrices for groups of users potential formed according to NOMA principles. For effective use of  NOMA, an optimization problem is formulated to maximize the sum rate of the network through  forming optimum groups under certain joint conditions, efficient power allocation, high quality of service for each weak and strong users, and high overall system performance. Such optimization problems are defined as  max-min fractional programs difficult to solve in practice. Therefore, a dynamic application for NOMA is introduced using two algorithms. First, a Radio frequency (RF)-aided dynamic algorithm is designed to form multiple groups, where  users exchange binary variables among them through an RF system to establish distance-based
weight edges, which are used as a metric for the grouping process. Second, a dynamic power allocation is proposed to determine the optimum power allocated to each group, while the users belonging to a certain group receive their traffic demands regardless of their classification as  weak or strong. The results show the convergence  of the proposed  dynamic application to the optimum solution, and  its high performance  in terms of sum rate, fairness, and energy efficiency compared to counterpart schemes. 
\end{abstract}

\IEEEpeerreviewmaketitle

\maketitle

\section{Introduction}
 
\IEEEPARstart{O}ptical wireless communication (OWC) is a well-known technology that  have been investigated to support various next-generation (6G and beyond) applications operating at  high data rates in a range of gigabit per second (Gbps).
In \cite{6011734}, light-emitting diodes (LEDs) were used in indoor environments to provide illumination and communication at high speeds. Interestingly, OWC systems have several advantages including the provision of high spectral and energy efficiency, low cost-infrastructure and low power consumption compared to Radio frequency (RF) networks. However, LED-based OWC systems have limited modulation speeds, and sending information at high communication speeds might affect the illumination function of the LED. Given that, infrared lasers, such as vertical-cavity
surface-emitting(VCSEL) lasers,  have  received enormous interest as a strong candidate in the next generation of OWC  due to their small size, long life time and high modulation speed. An IR laser provides usually uniform illumination within a small and confined coverage area in a range of a few square centimeters.  However, an array of IR lasers can be designed to expand the coverage area  to a few square meters. In an indoor environment, multiple IR laser arrays operating under eye safety regulations are needed to ensure uniform coverage and seamless transition for  multiple users on the floor \cite{9803253,9685357,9839259}. 

In OWC,  multi-user interference management is a crucial issue that plays a major role in determining the spectral efficiency of the network. Recently,  non-orthogonal multiple
access (NOMA) has been considered as a  promising approach to serve a high number of users simultaneously over the same resources, i.e., frequency or time slots, compared to orthogonal schemes \cite{8114722,8515272, 8030546,8361407}. Basically, the implementation of NOMA in wireless networks requires user-classification algorithms that form groups of strong and weak users  to exploit the power domain  by assigning a different power level to each user based on their channel gains. In \cite{7342274}, NOMA was applied in an OWC network considering realistic indoor optical channel conditions. In \cite{7572968}, the performance of NOMA was evaluated in OWC networks under the guarantee of high quality of service (QoS). 
Furthermore, the evaluation of bit-error-rate (BER) in NOMA-based OWC systems was addressed in \cite{7996769}, deriving a closed-form expression taking into account interference cancellation errors. In \cite{10949139}, the performance of NOMA was investigated in multiple-input multiple-output (MIMO) OWC systems, serving multiple users at high sum rates. It was shown that NOMA achieves higher data rates dictated by power allocation and the difference in channel gain between weak and strong users. 
\subsection{ Related Works}
The application  of NOMA to manage multi-user interference in wireless networks is subject to two main challenges. First, inter-group interference, which means that after  dividing the  users into  several groups, each user suffers  interference received due to transmission to the adjacent groups, resulting in significant degradation in the signal to interference and noise ratio (SINR). For instance,  
in \cite{7938651}, NOMA was applied in MIMO RF networks considering the case of multiple group generating interference among them. To eliminate this interference, it is proposed that the users of each group are served using a unique  transmit beamforming vector, which involves high cost in terms of complexity. 
In \cite{7442902}, a robust beamforming  was also designed for NOMA to serve multiple clusters of users in a multiple-input-single-output (MISO) RF setup under errors in channel estimation. OWC networks differ than RF networks in terms of density and the characteristics of the transmitted signal. Therefore,  the implementation of such beamforming techniques in OWC might lead to complex optimization problems, hence, orthogonal transmission between clusters of users was considered in \cite{10965784} and the majority of the NOMA-related work in OWC, compromising on overall system performance. Second issue in the application of NOMA especially in OWC networks is that each optical AP illuminates a small area referred to as attocell, and each group of users might experience low SNR due to inter-cell interference. In \cite{7792590},  NOMA was used to serve users in  a multi-cell OWC network, where a user-grouping algorithm is  designed  to determine user combinations taking into account their locations, while frequency reuse (FR) was applied  among the cells to avoid inter-cell interference. However, an indoor environment usually contains  a high number of optical APs, and therefore, managing each optical AP independently causes severe inter-cell interference. Moreover, using a frequency reuse technique in such high density scenarios is insufficient, and users might be subject to handover every few centimeters in dynamic OWC networks \cite{Qidan24}. 

At this point, a transmission scheme referred to as blind interference alignment (BIA) was proposed in \cite{5734870} to coordinate the transmission among RF transmitters and manage multi-user interference. Basically, the transmission block of BIA  allocates alignment blocks from the transmitters to each user following a certain methodology. In this case, each user  must have the ability to switch its reception modes  over its allocated alignment blocks following a predefined pattern. It was shown that a unique  antenna, namely reconfigurable antenna, is needed for providing a set of radiation patterns from the transmitters and enabling the application of BIA in RF networks. It is worth mentioning that the size of  transmission block of  BIA  increases with the number of transmitters and users, and therefore, the channel coherence time must be large enough to guarantee the delivery of such transmission blocks with minimum errors.
In \cite{8935164}, the methodology of BIA was adopted  to manage inter-group interference rather than muli-user interference in RF networks, while each group contains a pair of users served using NOMA. In \cite{8113509,9521837}, a reconfigurable optical receiver was proposed to enable the application of BIA-based transmission in OWC, where this receiver allows the user to switch its reception modes over time, providing linearly independent channel responses from a set of optical transmitters. In \cite{8636954,9064520,10279546}, new BIA schemes were proposed with the aim of enhancing its performance in OWC networks of high density. It is worth pointing out that  the application of  BIA in OWC is superior compared to orthogonal or transmit precoding schemes. It is because of BIA naturally coordinates the transmission among the transmitters, while each user can measure and cancel multi-user interference.
Given the features and limitations of BIA, this work adopts its application in NOMA-based OWC to overcome the limitations of NOMA in terms of inter-group interference and inter-cell interference, while exploiting the NOMA ability in providing multiple access service within a certain group.  

In NOMA transmission, transmitters send information using the power domain, giving  users the opportunity to  exploit  all the available resources. Therefore, suitable power allocation schemes are essential in NOMA. This issue has been investigated extensively in the literature in the context of RF and OWC systems. In \cite{7275086}, a novel gain ratio power allocation (GRPA) approach was proposed to allocate the power among the users considering  their  channel conditions. It was shown that the  GRPA technique achieves higher sum rate with some level of fairness compared to static power allocation algorithms. In \cite{7752879}, a  power control strategy was derived in a downlink OWC to maximize the sum rate of multiple users under the constraints of user-fairness and illumination. In \cite{10401944}, NOMA was implemented in a hybrid OWC/RF, where GRPA was used to distribute the power at a given user grouping. In \cite{8565925}, the performance of NOMA was enhanced in OWC through minimizing the transmit power per LED while achieving minimum data rate requirements for static users in the network. Interestingly, the formation of multiple groups in NOMA highly affects the power allocation techniques proposed in \cite{7275086,7752879, 10401944, 8565925}. Indeed, user grouping and power allocation in NOMA are defined as joint problems that must be solved to unlock the potential of  NOMA for use in the next generation of wireless communications. In \cite{7792590}, a user grouping approach was  proposed to form a number of groups in a multi-cell OWC network considering  the locations of the users, and accordingly, the sum rate of the network is maximized through performing power allocation within each independent optical cell. 
In \cite{9259258}, cooperative NOMA was adopted in a RF-aided single optical cell network where user-grouping and power allocation problems are solved to maximize the sum rate of the users belonging to a certain group. In \cite{10195155}, a permutation-based genetic
algorithm was proposed to optimize user
pairings and enhance the performance of hybrid NOMA and orthogonal transmission in an OWC network. Despite the detailed work on NOMA in the literature of RF and OWC networks, this paper defines  user-grouping, power allocation, user traffic demands, and overall system performance as joint problems that must be addressed  to promote NOMA as a strong candidate in the emerging 6G OWC.

\subsection{Main Contributions}
In this paper, an OWC network is considered using arrays of IR lasers as optical APs operating within a limited range of transmit power according to eye safety regulations. This network is known for its high density, where a large number of APs is required to ensure uniform coverage for multiple users. Therefore, we propose a BIA outer precoder to coordinate the transmission among the optical APs with the guarantee of interference alignment for multiple groups of users formed according to the NOMA methodology. Note that, in \cite{8935164}, a semi-Blind interference management scheme was proposed by integrating   BIA and NOMA in RF networks. To the best of our knowledge, this paper is the first to propose the integration of the two schemes in OWC.  Moreover, 
in contrast to related works, including \cite{8935164}, a joint rate-maximization optimization  problem is formulated to determine the optimal user grouping, efficiently distribute the power budget in the network, ensure the fulfillment of user traffic demands, and enhance overall system performance.
 
The main constitutions are listed as follows  

\begin{itemize}
\item An OWC system model is defined, which consists of a number of IR laser-based optical APs deployed to ensure uniform coverage for multiple-users. To overcome the limitations of NOMA, an outer precoder is designed  for AP coordination by following the methodology of BIA. This precoder has also the ability to align interference among groups of users that can be formed under certain conditions. Therefore, the application of NOMA can be applied successfully  within each group with relatively low noise, as a result of the effective management of both  inter-cell and inter-group interference.

\item  A rate-maximization optimization problem is formulated  subject to the constraints of unique elements within each group of users,  low power consumption, and  user traffic demand,  to ensure high performance in NOMA-based transmission. By solving this optimization problem, the  essential  conditions  mentioned above can be  guaranteed. However, the optimization problem is defined as a max-min fractional program, which is complex and difficult to solve in practical scenarios. It is  concave in nature and involves  coupled optimization variables.

\item  To provide practical solutions, an RF-aided dynamic algorithm is designed to form multiple groups taking into consideration the locations of the users, where a portion of the resources of an RF system deployed for uplink transmission \cite{9064520} is devoted for the strong and weak  users to exchange  binary variables and establish weak-strong user associations. Moreover,
 a dynamic power allocation algorithm is derived to maximize the sum rate of the users belonging to all the groups simultaneously, where different power levels are  allocated to different groups based on the demands of the users within each group. The application of the proposed algorithm guarantees that each user experiences high quality of service regardless of its classification as weak or strong, and that the power budget of the network is consumed efficiently, which in turn maximizes the overall performance of the  OWC network. 
\end{itemize}

The results demonstrate the high performance of the proposed dynamic application of NOMA in terms of data rate, energy efficiency, and fairness compared to BIA, traditional NOMA, and other benchmark schemes  introduced according to the literature of NOMA. Moreover, the dynamic user-grouping and power allocation algorithms provide practical and significant solutions close to the optimal ones  at low complexity. 

Other parts of  the  paper are organized as follows. In Section \ref{sec:system}, the system model of the laser-based OWC network is described. The transmission based on NOMA, inter-group interference management and the sum rate of the network are derived in Section \ref{sec:noma}.  The optimization problem  is formulated  in Section \ref{sec:ratem}. Section \ref{sec:dynamic} presents the dynamic user-grouping and power allocation algorithms in detail. Finally, Section \ref{sec:res} presents simulation results, and  Section \ref{sec:con}  provides concluding remarks and future directions. 
 
{\it Notation}.  The mathematical  notations invoked in this work are defined as follows.  Bold upper case and lower case letters denote matrices and vectors, respectively. $\mathbf{I}_M$ and $\mathbf{0}_M$ notations represent identity and zero matrices with $M\times M$ dimension, respectively. $\mathbf{0}_{M,N}$  denotes a $M\times N$ zero matrix. $[\,\,]^T$ and $[\,\,]^H$ indicate transpose and hermitian transpose operators, respectively. Finally, $\mathbb{E}$ is the statistical expectation, and $\mathrm{col}\{\}$ is the column operator that stacks the vectors considered in a column.

\section{System model}
\label{sec:system}
\begin{figure}[t]
\begin{center}\hspace*{0cm}
\includegraphics[width=1\linewidth]{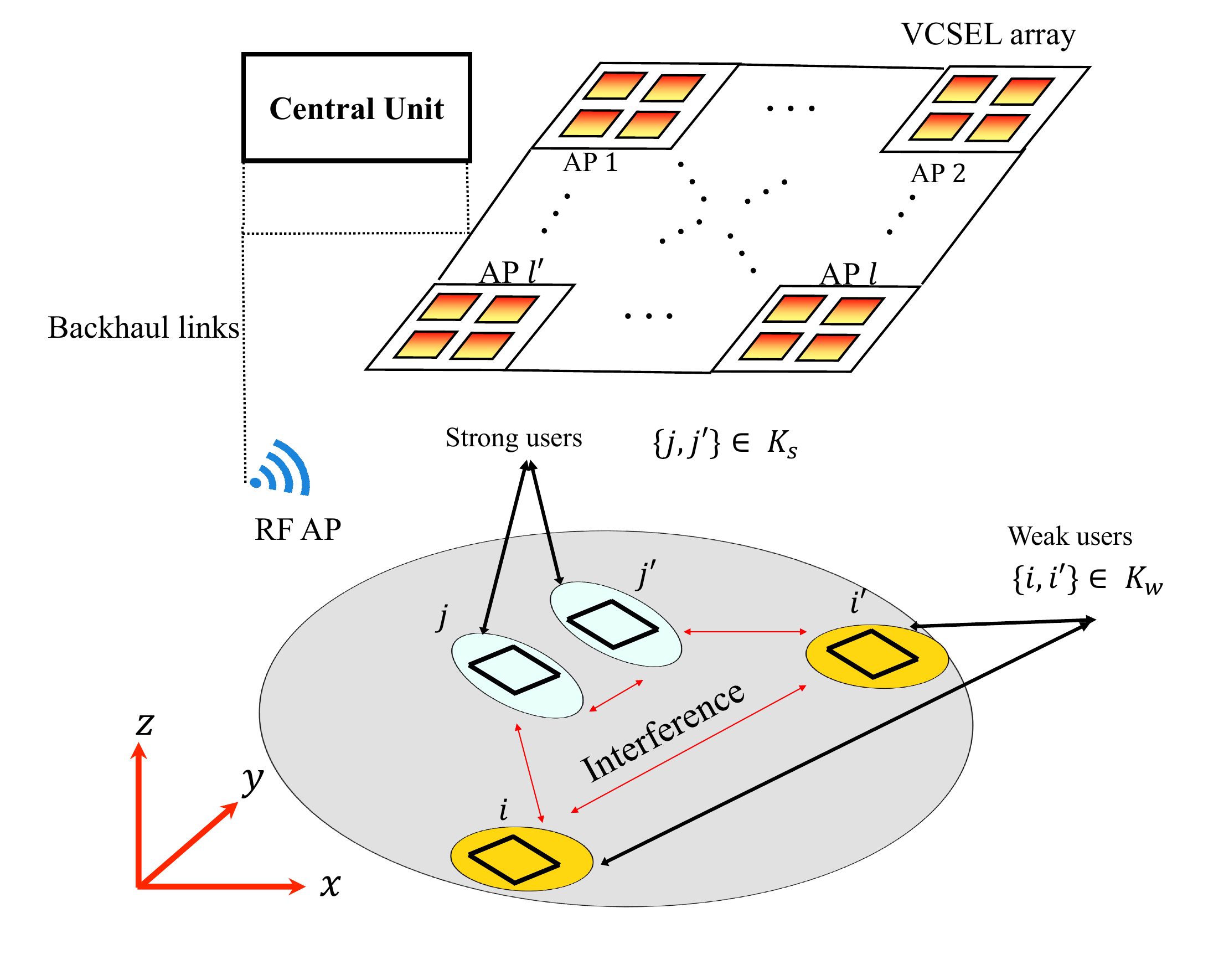}
\end{center}
\vspace{-2mm}
\caption{An OWC network using IR lasers as transmitters serving multiple users.}\label{Figmodel}
\vspace{-5mm}
\end{figure}

A downlink OWC network is considered as  shown in Fig.~\ref{Figmodel}. On the ceiling, a number of  optical APs  given by  $ L, l= \big\{1, \dots, L \big\} $, is deployed to provide  wireless converge for   
$ K $ users, where  $ k= \big\{1, \dots, K \big\} $.  In this work, all the $ L $ optical  APs  serve the $ K  $  users simultaneously where the transmission is coordinated using BIA \footnote{ The application of BIA over $ L $ optical APs results in full connectivity in the network \cite{8636954}. Note that, our system model can be managed as a multi-cell scenario where an optimum user-AP association vector must be determined and inter-cell interference must be managed using various network topology approaches. These issues are not within the scope of this work. Nevertheless, the formulation of the optimization problem and the practical solutions derived  in the paper can be easily extended to such multi-cell scenarios.}.
 Moreover, the users are classified as weak users denoted by $ K_w $,  where $ i= \big\{1, \dots, K_{w} \big\} $, and  as strong users denoted by $ K_s $,  where $ j= \big\{1, \dots, {K_s} \big\} $ \footnote{ The classification of the users is done according to the locations of the users, where the $ K_s  $  strong users receive high power from  all the  optical APs due to their locations at the center of the room, i.e., $  \mathrm{dist}\left(l,i\right) \leq d_{\mathrm{th}}, \forall l\in L, \forall i\in K_{s} $, where $ \mathrm{dist}\left(l,i\right) $ is the distance between user $ i $ and optical AP $ l $ and $ d_{\mathrm{th}} $ is the threshold distance  that results in a low received power. While the  $ K_w $ weak users located at the edges receive weak power from  the optical APs located at large distance \cite{9685357}. Without loss of
generality and to better understanding the derived schemes, a case of  $ K_{s}=K_{w} $ is considered in this work to form multiple groups each with a pair of users similarly to the majority of work in the NOMA literature, otherwise, $ K_{s}\neq K_{w} $, virtual weak or strong users can be assumed with zero data rates.}. Each user is equipped with a reconfigurable optical detector of  $ M, m= \big\{1, \dots, M \big\} $, photodiodes  to provide a wide field of view (FoV). This receiver also enables the user to switch its reception mode following a predefined pattern  to provide
linearly independent channel responses from $L$ optical APs \cite{8113509,9521837}.
In this context, the received signal at photodiode $ m $  of a generic  user $ k \in K $  from   $ L $ optical APs at time $ n $ can be expressed as 
\begin{equation}
y^{[k]}[n]=f\mih^{[k]}(m^{[k]}[n])^{T} \mix[n] + z^{[k]}[n],
\end{equation}
where $ f $ is the detector
responsivity, $ \mih^{[k]}(m^{[k]}[n]) \in \mathbb{R}_+^{L\times 1} $, $ m^{[k]}[n] $  is the channel state selected by photodiode $ m $ at time slot $ n $, $ \mix= [ x_{1}, x_{l}, \dots, x_{L} ]^{T} \in \mathbb{R}_+^{L\times 1}$  is the transmitted signal where $ x_{l}= \rho  \sqrt{P_{k}} u_{k,l}+ \rho\sum^{K}_{k'\neq k} \sqrt{P_{k'}} u_{k',l}+ \rho I^{Dc} $ is the transmitted signal from AP $ l $ towards user $ k $, where $ P_{k} $ is the power allocated to user $ k $, $ u_{k,l} $ is the desired message, $ I^{Dc}  $ is the direct current that ensures the non-negativity of the signal and $ \rho $ is the electric to optical power conversion factor. 
Moreover, $ \sum^{K}_{k'\neq k} \sqrt{P_{k'}} u_{k',l} $  is multi-user interference due to  transmission to other users, and  $ z^{[k]} $ is real valued additive white Gaussian noise with zero mean and variance  given by the sum of contributions from  shot noise,  thermal noise and the intensity noise of the laser. 
Finally, a WiFi AP is deployed  for uplink transmission, where users send through it important information for performing  various optimization problems with different objectives. All the APs in the network, optical and RF, are connected through a central unit (CU) that contains  information on the distribution of users and their need of resources. It is worth mentioning that the need for CSI at the transmitters is subject to the use of the  multiple access technique implemented  to serve the active  users in the network.
 \begin{table*}[h]
\centering
\caption{Transmitted signal for $L = 2$, $G=2$, $ K_g=\{i,j\}, K_{g'}=\{i',j'\}, g'\neq g $}
\label{tabla_signal}
\begin{tabular}{c|c|c|}
\cline{2-3}
                                  & AP 1 & AP 2 \\ \hline
\multicolumn{1}{|l|}{Time slot 1} &  $(u_{i,1}+ u_{j,1}) + (u_{i',1}+ u_{j',1})$    &  $(u_{i,2}+ u_{j,2}) +( u_{i',2}+ u_{j',2})$\\ \hline
\multicolumn{1}{|l|}{Time slot 2} & $u_{i,1}+ u_{j,1}$     & $u_{i,2}+ u_{j,2}$\\ \hline
\multicolumn{1}{|l|}{Time slot 3} & $u_{i',1}+ u_{j',1}$    & $u_{i',2}+ u_{j',2}$    \\ \hline
\end{tabular}
\end{table*}
\subsection{ Optical transmitters}  
In this work, the VCSEL is  used as a transmitter operating under eye safety regulations due to its high modulation speed, and  each optical AP is defined as an  array of  $ L_{v} \times L_{v} $ VCSELs to expand its confined coverage area. Interestingly, the VCSEL has a Gaussian beam profile, and  parameters  such as  
 the beam waist $ W_{0} $,  wavelength $ \lambda $ and  beam travelling distance $ d_0 $  determine the power distribution of the VCSEL. In this context,  the beam radius of the VCSEL  at the receiving plane located at distance $ d_0 $  is given by 
\begin{equation}
W({d_0})=W_{0} \left( 1+ \left(\frac{d_0}{d_{Ra}}\right)^{2}\right)^{1/2},
\end{equation}
where $ d_{Ra} $ is the Rayleigh range expressed as $ d_{Ra}= \frac{\pi W^{2}_{0} n }{ \lambda},$ where $ n $ is the refractive index of the air, i.e., $ n=1 $. Moreover, the spatial distribution of the intensity of the VCSEL over the transverse plane at the same distance $ d_0 $ is given by 
\begin{equation}
I_{tr}(r,{d_0}) = \frac{2 P_{tr}}{\pi W^{2}({d_0})}~ \mathrm{exp}\left(-\frac{2 r^{2}}{W^{2}({d_0})}\right),
\end{equation}
where $ r $ is the radial distance from  the beam spot center and  $ P_{tr} $ is the optical power per the beam. Note that, the detection area of each photodiode within the reconfigurable detector  can be given by $ A_m = \frac{A_{rec}}{M} $ with radius $ r_{m} $, where  $ A_{rec} $ is the whole detection area of the optical detector. Therefore, 
the  optical power received at photodiode $ m $ of user $ k $ from  a transmitter is given by 
\begin{multline}
P_{m}= 
\int_{0}^{r_{m} } I_{tr}(r,d_0) 2\pi r dr =  P_{tr}\left[1- \mathrm{exp} \left(\frac{ -2{r^{2}_{m}} }{ {W^{2}({d_0})}}\right)\right],
\end{multline}
considering a scenario of the  photodiode $ m $ of user $ k $ is aligned with the transmitter. It is worth mentioning that each photodiode of the reconfigurable detector points to a distinct direction determined by elevation and azimuth angles, $\theta_{k,m}$ and $\alpha_{k,m}$, respectively, on the x-y plane. Thus, the normal vector of  photodiode $m$ of user $k$ is 
\begin{equation}
\begin{split}
\hat{\mathbf{n}}_{k,m}=&
\left[ \sin\left(\theta_{k,m} \right)\cos\left(\alpha_{k,m}\right), \,\, \right .\\
 &\phantom{\{} \left . \sin\left(\theta_{k,m} \right)\sin\left(\alpha_{k,m}\right), \,\, \cos\left(\theta_{k,m} \right) \right],
\end{split}
\end{equation} 
In different scenarios, the irradiance angle of the VCSEL in regrade to user $ k $ is  determined by
$\phi_{tr}^{[k]}=\cos^{-1} \left(\frac{\hat{\mathbf{n}}_{tr} \cdot {\mathbf{d}}}{d} \right)$, considering that  $ {\mathbf{d}}= \mathbf{P}_{tr}- \mathbf{P}_k $ is  the distance vector given by the locations of the transmitter and user, $ \mathbf{P}_{tr} $ and  $ \mathbf{P}_k $, respectively,  $ d= \| \mathbf{d}\|$. At this point, $ r = d \sin \phi_{tr}^{[k]} $,  and  $ d_0 $  is replaced by $ d \cos \phi_{tr}^{[k]} $. Hence , the spatial distribution of the intensity of the VCSEL is defined as a function of the distance $ d $  and irradiance angle $ \phi $, i.e., $ I_{tr}({d}, \phi) $. In this context, the optical power received at photodiode $ m $ of user $ k $ from  the VCSEL is given by 
\begin{multline}
P_{m}= I_{tr}({d},\phi) A_{pd} G_{m} \cos (\psi_{tr}^{k} (m)) $rect$ \left(\frac{\psi_{tr}^{k} (m)}{\Psi_{F}}\right)\\
= \frac{2  \cos (\psi_{tr}^{k} (m)) P_{tr} A_{pd} G_{m}}{\pi W^{2}(d \cos \phi_{tr}^{[k]})}\\ \mathrm{exp}\left(-\frac{2 d^{2} \sin^{2} \phi_{tr}^{[k]}}{W^{2}(d \cos \phi_{tr}^{[k]})}\right) $rect$ \bigg(\frac{\psi_{tr}^{k} (m)}{\Psi}\bigg),
\end{multline}
where $  A_{pd} $ and $ G_{m} $ is the detection area  and gain of the photodiode, respectively. Moreover, $ \psi_{tr}^{k} (m) $ is the angle between  the normal vector $ \hat{\mathbf{n}}_{k,m} $ and the distance vector $ {\mathbf{d}} $, i.e.,  $\psi_{tr}^{k} (m)=\cos^{-1} \left(\frac{\hat{\mathbf{n}}_{k,m}  \cdot {\mathbf{-d}}}{d} \right)$. Finally, $ \Psi_{F} $ is  the  field of view  where $ $rect$ \bigg(\frac{\psi_{tr}^{k} (m)}{\Psi}\bigg)=1 $  if $ 0 \leq \psi_{tr}^{k}  \leq\Psi_{F} $, or it equals zero otherwise. 
\subsection{Optical constraints}  
The use of VCSEL maximizes the communication speed of OWC due to its high modulation speed. However, the transmit optical  power $ P_{tr} $ is limited for  eye safety regulations. To calculate the maximum transmit power allowed per VCSEL $  P_{tr,\max}  $, we define some  notations as follows.  The aperture diameter of the cornea  is $ d_c $,  the exposure level of the cornea to the transmit power is  $ E_{e} $, the maximum exposure level  is $  E_{e,\max}(t_e) $ as a function of the exposure duration $ t_e $ and the hazard distance is $ d_h $ at which eye injures might happen. In this context,  $ E_{e} $ at the hazard distance $ d_h $ can be calculated according to \cite{9803253} as  
 
\begin{multline}
E_{e}(d_h)= \frac{1}{\pi (\frac{d_c}{2})^{2}}
\int_{0}^{d_c/2 } I(r,d_h) 2\pi r dr \\ =\frac{1}{\pi (\frac{d_c}{2})^{2}} \int_{0}^{d_c/2 } \frac{2 P_{tr}}{\pi W^{2}({d_h})}~ \mathrm{exp}\left(-\frac{2 r^{2}}{W^{2}({d_h})}\right)2\pi r dr\\=
\frac{P_{tr}}{\pi (\frac{d_c}{2})^{2}} \bigg(1-\mathrm{exp}\left(-\frac{d_{c}^{2}}{2 W^{2}({d_h})}\right)\bigg).
\end{multline} 
Interestingly, the transmit power of the VCSLE is safe for the human eye if $ E_{e}(d_h)\leq E_{e,\max}(t_e) $, which means that the use of the VCSEL is safe regardless of the location of the user as in the most hazard location, the transmit power does not cause any harm to the human eye. Therefore, the maximum permissible power per beam is determined as   
\begin{equation}
  P_{tr,\max}= \frac{\pi}{4} d_{c}^{2} E_{e,\max}(t_e) \bigg( 1- \mathrm{exp} \bigg( \frac{d_{c}^{2}}{2 W^{2}({d_h})}\bigg) \bigg)^{-1}.
\end{equation} 
More details on the calculation of $E_{e,\max}(t_e)  $ can be found in \cite{9803253}. Given the eye  safety regulations, the deriving current \footnote{A fixed DC-biasing is considered to focus our attention  on optimizing power allocation among the messages intended to different users as adaptive signal scaling and DC biasing  are critical in OWC that require independent studies \cite{Wang:14}. } of the VCSEL $ I_{v}^{Dc} $ must be within a certain range of current $ [I_{L}, I_{H}] $, 
ensuring that the VCSEL works in the linear region and that the corresponding deriving power $ [P_{L}, P_{H}] $ guarantees eye safety. As a consequence, the peak amplitude of the modulated signal  applied to the VCSEL, which  might contain information intended  to $ K $ generic users, is denoted by $ \eta_{v} $, and it must be subject to the constraint $ \eta_{v} \leq \min (I_{v}^{Dc}-I_{L}, I_{H}-I_{v}^{Dc})$.
It is worth pointing out that each AP $ l \in L $ is composed of $ L_{v}\times L_{v} $ VCSEls, and therefore,  $ P_{L} \leq (P_{l})/(L_{v}\times L_{v}) \leq  P_{H} $, where  $ P_{l} $ is  the transmit power of the AP, ensuring that each VCSEL in the array of the AP emits the same power, i.e., $  P_{tr,\max}=(P_{l})/(L_{v}\times L_{v}) $. 

 \section{NOMA-based Transmission}
\label{sec:noma}
For an OWC network as shown in Fig.~\ref{Figmodel}, we first design an outer precoder to coordinate transmission among $L$ optical APs and to align interference across $G$ groups formed based on channel gain differences, where  each group $ g \in G $ consists of $ K_g=\{i,j\} $ users, classified as weak and strong users, $ i $  and $ j $, respectively. Then, NOMA is considered within each group to manage intra-group interference, deriving the sum rate of the network.

\subsection{Outer precoder design}
An outer precoder is designed following the methodology of BIA proposed first in \cite{5734870,8935164}. 
 
Basically, BIA generates a transmission block over which alignment blocks are allocated to each group. Each alignment block is composed of $ L $ time slots,  and the users of a certain group switch  their reception modes over each of their alignment blocks following the same distinctive pattern to receive useful information. The construction of the BIA transmission block can be easily determined considering the number of APs and groups. In this case, it must contains $ (L-1)^{G}+G(L-1)^{G-1} $ time slots divided into two sub-blocks, where sub-block 1 contains $ (L-1)^{G} $ time slots and sub-block 2 contains $ G(L-1)^{G-1} $ time slots. At this point, the number of alignment blocks allocated to each groups is $\zeta=\{1, \dots, (L-1)^{G-1}\}$ distributed over the transmission block with two conditions as follows 

\begin{itemize}
\item \textbf{Ensuring coordination among $ L $ APs and the decodability of each symbol.} Each symbol transmitted to a certain group denoted by $\mathbf{u}_{\zeta}^{[g]}$ over the $\zeta$-th alignment block must contains $ L K_{g} $ messages from $ L $ APs, i.e., $ \mathbf{u}_{\zeta}^{[g]}=[u_{i,1}+ u_{j,1}, \dots, u_{i,L}+ u_{j,L} ]^{T} $, where $ u_{i,l}$ and $ u_{j,l} $ are the messages transmitted by AP $ l $ to the weak and strong users forming group $ g $, respectively. Thus, the users of each group must switch their reception modes among $L$ modes, i.e., a predefined pattern, during the transmission of the $\zeta-$th alignment block to achieve  $  L K_{g} $ degrees of freedom (DoF) in $\mathbf{u}_{\zeta}^{[g]}$.   

\item \textbf{ Inter-group interference alignment.} BIA can align the interference among multiple groups if the users $ K_{g'} $ belonging to  group  $ g'\in G $, $ g^{'} \neq g $  can measure  the interference due to the transmission of the symbol  $\mathbf{u}_{\zeta}^{[g]}$  over the $\zeta$-th alignment block  allocated  to the $ K_{g} $ users belonging to group $ g\in G $. It can be achieved  by ensuring the reception  modes  of the  $ K_{g'} $ users remain the same while  the $ K_{g} $ users change their reception  modes to receive useful information transmitted over an alignment block. Given that, the interference can be aligned in at least one dimension less compared to the desired information.
\end{itemize}
For illustrative purpose, a use  case of $ L=2 $ optical APs serving $ G=2 $ groups, each group with $ K_g=2 $ users, is shown in Table~\ref{tabla_signal}, where the transmission block comprises  $ (2-1)^{2}+2(2-1)^{2-1} $ time slots, and $ \zeta= (2-1)^{2-1} $ alignment blocks are allocated to each group. Note that, the $ K_g $ users are served over the time slots \{1,2\} constituting the only one alignment block allocated to group $ g $ in this scenario, while  the $ K_{g'} $ users are served over the time slots \{1,3\}. Moreover, the users of group $ g $ use the third time slot to measure  and subtract the interference received over the first time slot, while the users of group $ g' $ use the second time slot to measure and cancel the interference, due to the fact that orthogonal transmission is occurred over the second and third time slots. More mathematical details can be found in \cite{5734870}. 

At this point, the transmitted signal for the general case  can be expressed as 
\begin{equation}
\mathbf{X}= \sum^{G}_{g=1} \mathbf{B}^{[g]} \bigg( \sqrt{P_w} \mathbf{u}_{i,T}^{[g]}+\sqrt{P_s}\mathbf{u}_{j,T}^{[g]}\bigg)+\rho I^{Dc} \times \mathbf{1}_{L\times1}, 
\end{equation}
where $ {P_w} $ and $ {P_s} $ is the power allocated to the weak and strong users, respectively. Moreover, $ \mathbf {X}= \text{col} \{\mathbf{x}(\kappa)\}_{\kappa=1}^{\mathcal{V}_L} $, where $ {\mathcal{V}_L}$ is the length of  the transmission block designed according to the principles of BIA,   $ \mathbf{B}^{[g]}$ is the precoding matrix  used by the $ K_g $  users belonging to group $ g $, and $ \mathbf{u}_{k,T}^{[g]} =\text{col} \bigg\{ \mathbf{u}_{k,\zeta}^{[g]} \bigg\}_{\zeta=1}^{(L-1)^{G-1}} $, where $ \mathbf{u}_{k,\zeta}^{[g]}=[u_{k,1}, \dots, u_{k,L}]^{T} $ contains the  symbols transmitted to user $ k $ over each alignment block allocated to group  $ g $, $ g\in G$, during the entire transmission block. Interestingly,  once the transmission block is built and multiple  alignment blocks are allocated to each group  with the two conditions above,  the precoding matrix $ \mathbf{B}^{[g]}$ used by the users of the same  group  can be easily determined, where each alignment block $ {\zeta} $ represents a column in the precoding matrix, stacking $ L \times L $ identity matrices in the rows  corresponding to the time slots that form each  alignment block.  After designing the precoding matrices from $L$ APs to  $ G $ groups, each user $ i\in K_{g} $  is  subject only to intra-group interference from its paired user $  j\in K_{g} $, $ K_{g}=\{i,j\} $, which can be canceled using NOMA as below. 
\subsection{NOMA}
The received signal at  weak user $ i \in K_{g}  $  can now be written as
 \begin{multline}
\tilde{\mathbf{y}}_{g,i}= \rho f\bigg({\mathbf{H}}^{[g,i]} \sqrt{P_w} \mathbf{u}_{i,\zeta}^{[g]}+ \underbrace{+ {\mathbf{H}}^{[g,j]} \sqrt{P_s}\mathbf{u}_{j,\zeta}^{[g]}}_{\text{intra-group interference}}\bigg)+ \tilde{\mathbf{z}}^{[g,i]}.
\end{multline}
Note that, the received signal of the strong user $ j \in K_{g} $ can be easily derived following the same fashion. It is worth pointing out that the  power levels $ {P_w} $ and   $ {P_s} $ must be determined carefully, ${P_w} > {P_s}$, giving each user the ability to decode its message \cite{7342274}. Therefore, the achievable rate  of a generic user $ k $ in a certain group $ g $ considering  the use of the outer precoder and NOMA  can be expressed as follows

\begin{equation}
\label{ratep}
R_{g,k}=  b_{ab} \times\mathbb{E}\left[ \log \det\left( \mathbf{I} +  \gamma_{g,k} \mathbf{H^{[g,k]}}\mathbf{H^{[g,k]^{H}}}  \mathbf{R_{\tilde{z}}}^{-1}\right)\right],
\end{equation}
where   $ \gamma_{g,k} $ is the SINR of a generic user $ k $ taking into account
the distinct features of the optical signal. It is given by  $ \gamma_{g,i}=\frac{c\rho^{2} {f}^{2} {P_{w}}}{c\rho^{2} {f}^{2} {P_{s}}+{\sigma_{z}}^{2}}
 $ and $ \gamma_{g,j}=\frac{c\rho^{2} {f}^{2} {P_{s}}}{{\sigma_{z}}^{2}}
 $ for the weak and strong users, respectively, where $ c=1/2 \pi e $ considering $ e $  as the Euler number. Moreover, $ b_{ab} $ is the ratio of alignment blocks allocated to each group due to the use of the outer precoder, it is given by $ \frac{1}{L+G-1} $, and $ \mathbf{R_{\tilde{z}}} $ is  the covariance matrix that results from  subtracting the information sent to $ G-1 $ groups.
\section{Sum rate maximization in NOMA-based OWC}
\label{sec:ratem}
To enhance the performance of NOMA-based transmission in dense OWC networks, we formulate here an optimization problem that jointly optimizes user grouping and power allocation to maximize overall system performance while satisfying the traffic demands of all users, regardless of their classification as  $ K_w $ or $ K_s $.

Let us  first define an association variable as  $ \mathit{x^{[i,j]}}$, $ \{i,j\}\in K $, in order to  pair each weak user $ i $ with a strong user $ j $. Therefore, it equals to  
 
\begin{equation}
\label{ass1} 
\mathit{x^{[i,j]}} = 
\begin {cases}
1 & $,     if weak user $  i$ is paired with strong user $ j $, $ \\
0 &, $otherwise.$ 
\end{cases}
\end{equation}
In this context, the sum rate of a pair of weak and strong users can be expressed as follows 

\begin{equation}
R_{g,K_g} (P_{w},P_{s},x)= x^{[i,j]} \bigg( R_{g,i}({P_{w}},P_{s})+ R_{g,j}({P_{s}})\bigg), 
\end{equation}
where $ K_g=\{i,j\} $ contains $ i $ and $ j $ users that form the potential group $ g $. Note that, according to \eqref{ass1}, $ R_{g,K_g} (P_{w},P_{s},x)=0 $  if weak user $ i $ is not paired with strong user $ j $, i.e., $ x^{[i,j]}=0 $, otherwise, $ x^{[i,j]}=1 $, and   $ R_{g,K_g} (P_{w},P_{s},x) $ can be determined easily from \eqref{ratep} after the application of BIA-NOMA for interference management, $ g'\neq g $, $K_{g} \neq K_{g}$, and $ i\neq j $. An appropriate optimization problem can then be formulated to maximize the overall network performance as follow   
 
\begin{subequations}
\label{op0}
\begin{align} 
\max_{\bm{p}} \quad & \bigg\{\min_{\bm{\{i,j\}\in K}} \bigg\{ x^{[i,j]} \bigg( R_{i}({P_{w}},P_{s})+ R_{j}({P_{s}})\bigg) \bigg\}\bigg\}\\
\textrm{s.t.}  \quad  & \sum\limits_{i\in K_w} \sum\limits_{j\in K_s}  (P_{w}+P_{s}) \leq P_{\max},~~~\forall i,j\in K \\
\quad  & \sum\limits_{i\in K_w} \mathit{x^{[i,j]}}=1, ~~~~~~~~~~~~~~~~~ \forall j\in K_{s}\\
\quad  & \sum\limits_{j\in K_s} \mathit{x^{[i,j]}}=1, ~~~~~~~~~~~~~~~~~ \forall i\in K_{w}\\
\quad  &  P_{s} \leq P^{T}_{s}, \label{op0_1} \\ 
\quad  &  P_{w} > P^{T}_{s}, \label{op0_2} \\ 
\quad & R_{\min,k} \leq R_{k}(P) \leq R_{\max,k}, ~~~\forall k\in K  \label{op0_3}  \\ 
\quad  & x^{[i,j]} \in \{0,1\}, P_{s} \geq 0, \{ K_{w}\cup K_{s}\}=K,
\end{align} 
\end{subequations} 

Physically, the first constraint in \eqref{op0} ensures that the power consumption in the network is equal or less than  the maximum  transmit power $ P_{\max}= \sum^{L}_{l=1} P_{l} $, where $ P_{l} $ is determined according to the eye safety regulations as in the system model. The second and third constraints guarantee that every group contains a unique pair of weak and strong users, i.e., $ \{i,j\}\in K_{g}, K_{g} \cap K_{g'}=\emptyset, g \neq g' $. Moreover, the fourth constraint in \eqref{op0_1} is imposed to limit the power allocated to the strong user such that the weak user manages the desired information of the strong user as noise while decoding its information \cite{7572968}. The power  constraint in  \eqref{op0_2} guarantees that the weak user in a group receives higher power than the strong user, and therefore, users performing SIC can decode the information  successfully in descending order. The constraints \eqref{op0_1} and \eqref{op0_2} also ease measuring the interference that results from the transmission to other groups \cite{7437435}. Furthermore, the data rate constraint  in  \eqref{op0_3} ensures high quality of service for all the $ K $ users regardless of their classification as strong or weak users. It is worth mentioning that the data rate of each user goes towards the maximum value $ R_{\max,k} $ if and only if other users receive their demands and the power consumption is less than $ P_{max} $. Otherwise, the data rate of each user goes towards the minimum data rate $ R_{\min,k} $, ensuring that all the users evenly experience high quality of service. The last constraint gives the fact that the strong user receives  a non-zero power and the feasible region of the optimization problem formulated based on the principles of NOMA.

The optimization problem in \eqref{op0}  is classified as a  max-min fractional program that has high complexity with a concave objective function \cite{opop,2009convex}. Moreover, the coupling of power allocation  and user grouping makes the optimization  problem  infeasible to solve especially in scenarios that require practical solutions with low computational time. Therefore, we propose to divide the main problem into two sub-problems that can be solved separately to reduce the complexity. In particular, weak and strong users can form multiple groups based on the distance rather than the achievable data rates, where it was demonstrated in \cite{7572968} that  the application of NOMA in wireless communications among two users becomes more successful if those users considerably differ in their channel gains. Then, power allocation can be performed among the groups formed earlier to maximize the overall sum rate of the NOMA-based OWC network, while ensuring that the users in each group receive their traffic demands. It is worth mentioning that the reduction  in complexity might come at the  cost of providing sub-optimal solutions.     
           
\section{ Dynamic User grouping and Power Allocation Algorithms for NOMA}
\label{sec:dynamic}

 We design an RF-aided dynamic algorithm to perform user grouping based on the distance between users that can potentially be grouped, ensuring a high channel gain difference between them. Subsequently, a closed-form optimization problem is formulated for power allocation among the formed groups.
This power allocation problem can be solved using an efficient dynamic algorithm that ensures  fulfilling the traffic demands of the users belonging to all the group, which in turn minimizes power waste and maximizes the total sum rate of the network. In the following, theses dynamic algorithms for user grouping and power allocation are reported in detail.    

\subsection{Dynamic user-groping algorithm for NOMA}
A distance-based weight  denoted as $ w(\mathrm{dist}) $ is considered to determine  optimal weak user-strong user combinations. It is worth mentioning that in the system model considered, an RF system is deployed to provide uplink transmission in the network. In this context, some of the RF system resources $ (1-e_{ \mathrm{rf}}) $, where $ e_{\mathrm{rf}} $ is a real value between
0 and 1,  are dedicated for the users to exchange binary variables among them. In \cite{7572968}, it was reported that NOMA provides a high sum rate in a group, whose users, weak and strong users, differ considerably in the channel gain. Given that, the users in the network can be considered as vertices distributed on the receiving plane. Then, each  strong vertex  sends  a binary variable  through the RF system to all the weak  vertices to establish  distance-based weight edges. For instance, the distance-based weight of an edge $ w(\mathrm{dist}) ({j\rightarrow i})$ between weak user $ i $ and strong user $ j $ can be given as follows 
\begin{equation}
\label{wight}
w(\mathrm{dist} ({j\rightarrow i}))=\sqrt{(x_{i}-x_{j})^{2}+(y_{i}-y_{j})^{2}},
\end{equation}
where $ (x_{i}, y_{i}) $ and $ (x_{j}, y_{j}) $ are the coordinates of weak user $ i $ and strong user $ j $, respectively. Note that, the distance-based weight of edge $ {j\rightarrow i} $ in \eqref{wight} increases with the distance between weak user $ i\in K_{w} $ and strong user $ j \in K_{s} $. Interestingly, the channel gain of the user is dictated by its location from the optical transmitters. Therefore, the strong users are characterized by their high channel gains due to their locations at the center of the receiving plane, while the weak users  located away from the optical transmitters receive a low channel gain. In this context,  weak user $ i \in K_{w} $ must be paired with strong user $ j \in K_{j} $ if and only if the edge $ {j\rightarrow i} $ has the highest weight compared to other edges between strong user $ j $ and other weak users, $  {j\rightarrow i'}  $, where  $ i'\neq i $, and $ i' \in K_{w} $. That is, each group can be formed as follows
\begin{equation}
\label{pro-c}
\begin{aligned}
g^{*}(i,j)=\arg\max_{x} \quad & \sum\limits_{i\in K_{w}} \sum\limits_{j\in K_{s}}\mathit{x^{[i,j]}} w(\mathrm{dist} ({j\rightarrow i})), \\
\end{aligned}
\end{equation}
where  $ \sum\limits_{j\in K_s} \mathit{x^{[i,j]}}=1, \forall i\in K_w $, $ \sum\limits_{j\in K_s} \mathit{x^{[i,j]}}=1, ~~~ \forall i\in K_w\ $ and $ \mathit{x^{[i,j]}}\in \{0,1\}, K_{w}\cup K_{s}=K $, ensuring that each group $ g $ is composed of unique elements according to the constraints in \eqref{op0}. 
This optimization problem can be solved through simply searching for optimal weak user-strong user combinations, forming $ G $ groups, each group with $ K_{g}=\{i,j\} $ distinct users, i.e.,
\begin{multline}
\label{uniq}
~\begin{matrix}
{K}_{g}= i \cup j, {K}_{g^{'}}= i' \cup j', {K}_{g} \cap {K}_{g^{'}}=\emptyset, ~\\ (i \neq i^{'}, j \neq j^{'}, g \neq g^{'}), 
\big\{{i,i^{'}}\big\} \in {K}_{w}, \big\{{j,j^{'}}\big\} \in {K}_{s}, ~\\ \big\{{g,g^{'}}\big\} \in G,{K}_{w}\cup {K}_{s}= K.
\end{matrix}
\end{multline}
\subsection{Power allocation}
Once each weak user is paired with a strong user, the  problem in hand is to perform power allocation  among multiple  groups  formed at the earlier stage and under the constraints of power budget and high quality of service. Therefore, the optimization problem in \eqref{op0} can be rewritten as follows 
\begin{subequations}
\label{op6}
\begin{align} 
\max_{\bm{p}} \quad & \bigg\{\min_{\bm{g\in G}} R_{g,K_{g}}(P_{w},P_{s})\bigg\}\\
\textrm{s.t.}  \quad  & \sum\limits_{i\in K_w} \sum\limits_{j\in K_s}  (P_{w}+P_{s}) \leq P_{\max}, ~~~\forall i,j\in K \\
\quad  &  P_{s} \leq P^{T}_{s},\\
\quad  &  P_{w} > P^{T}_{s},\\
\quad & R_{\min,k} \leq R_{g,k}(P) \leq R_{\max,k}, ~~~\forall k\in K \\
\quad  & P_{s} \geq 0, \{K_{w}\cup K_{s}\}=K, g \in G,
\end{align} 
\end{subequations}
where $ R_{g,K_{g}}(P_{w},P_{s}) $ is the sum rate of the group comprising $ K_{g} $ users. To solve \eqref{op6}, we introduce a dynamic power allocation algorithm to provide significant sum rate maximization at lower cost compared to solving the complex optimization  problem  in \eqref{op0}. It is a practical algorithm that runs for the $ K_g $ users of each group to achieve the highest sum rates while ensuring that the users belonging to the other groups experience high data rates. In particular, the power constraint in \eqref{op6} is divided by a number  $ {T} $, where $ t=\{1,\dots, t,\dots, T\} $, to have levels of the power constraint. That is 

\begin{equation}
\label{powercont}
P_{max,t}=\left\{\frac{t P_{max}}{T}: t=\{1,2, \dots, T\}\right\}.
\end{equation}     
Assuming at the beginning of the proposed algorithm that no user is served,  and therefore the sum rate of the network $ R_{n,G}(P_{w},P_{s}) $ under the power constraint $ P_{max,t}  $ can be expressed as follows 
\begin{multline}
R_{n,G}(P_{w},P_{s})=0, \textrm{s.t.}  ~~ \sum\limits_{i\in K_w} \sum\limits_{j\in K_s} (P_{w}+P_{s}) \leq P_{max,t},\\ P_{w}=0, P_{s}=0, \forall g\in G, \forall K_{g}\in K, \forall t\in T,   
 \end{multline}
Subsequently, the sum rate of the $ K_{g} $ users belonging to group $ g $ selected randomly in this stage must be determined at each power constraint with the condition of dedicating  power for the other users belonging to the groups considered prior to group $ g $. Therefore, under the power constraint $ P_{max,t} $, the optimization problem can be simply  expressed as 

\begin{subequations}
\label{op8}
\begin{align} 
\max_{\bm{p}} \quad &  R_{g,K_{g}}(P_{w},P_{s})\\
\textrm{s.t.}  \quad  &   (P_{w}+P_{s}) \leq P_{max,t},~~~ ~~~ ~~~ ~~~\forall t \in T \label{op8_1}  \\ 
\quad  &  P_{s} \leq P^{T}_{s}, \label{op8_1b} \\  
\quad  &  P_{w} > P^{T}_{s}, \label{op8_2} \\ 
\quad & R_{\min,k} \leq R_{g,k}(P) \leq R_{\max,k}, ~~~\forall k\in K_{g} \label{op8_3} \\ 
\quad  & P_{s} \geq 0, \{i,j\}\in K_{g}, K_{g}\in K ,g \in G. 
\end{align} 
\end{subequations}  
Note that, the achievable data rate of the weak user is a function of  the power values allocated  to the strong and weak users due to the interference term in \eqref{ratep}. Given that, the assumption of serving the strong users at  the highest possible power $ {P^{T}_{s}}$ is considered, and  an upper bound of multi-user interference received  at the weak user can be derived, which means that the interference is a constant \cite{7437435}.

\textbf {Proposition}:{\it A parametric approach at a given parameter \cite{opop} can be invoked to solve the optimization problem in \eqref{op8}. That is, the objective function can be transformed into
\begin{equation}
\label{di}
\mathcal{L}=\max_{\bm{p}}  \bigg\{ R_{g,K_{g}}(P_{w},P_{s})-\theta \bigg\},
\end{equation}
where $ \theta=\xi_{g} (P_{w}+P_{s}) $, considering that $ \xi_{g} $ is a non-negative parameter given by $ R_{g,K_{g}}(P_{w},P_{s})/ P^{T}_{g}  $, where $ P^{T}_{g} $ is the total power consumed by the pair of users, $\{ i,j\}\in K_{g} $.}
\begin{proof}
See the Appendix 
\end{proof}
Interestingly, if each of the $ K_g$ users receive  the maximum data rate  $ R_{\max,k} $ and the power consumption $ P_{\max,g} >  P_{max,t}$, it means that the $ (K'-K_{g}) $ users, where $ K' $ is the number of the users belonging to the groups considered until this stage,  receive 0 power under the power constraint $P_{max,t}  $. Therefore, the data rate of each of the $ K_g$ users must decrease towards the minimum data rate $ R_{\min,k} $, such that the power constraint $ P_{max,t}  $ can be  enough to support at least $ K_g $ users.
Otherwise, if  the power consumption $ P_{\max,g}$ is still higher than $ P_{max,t}  $, then, there is  no feasible solution for the network to support $  (K')  $ users under that power constraint. Therefor, the power is slightly relaxed to  $ P_{max,t+1}  $, and the same procedure is followed considering the new power constraint. 

Let us consider the other scenario in which the power  consumption  $ P_{\max,g} \leq P_{max,t}$, and the $ K_g $ users receive data rates within the specified range $ R_{\min,k} \leq R_{g,k}(P) \leq R_{\max,k} $.   In this case,  the fractional power resulting from $( P_{max,t}-P_{\max,g})  $ is used to support $ (K'-K_{g}) $ users, each $ K_g $ belongs to a certain group. Therefore, the overall maximum sum rate of the network $  R_{n,G'}(P_{max,t}) $, where $ G' $ is the number of the total groups considered until this stage, under the power constraint $ P_{max,t} $ can be determined according  to a recursive equation as follows  
\begin{multline}
R_{n,G'}(P_{max,t})= \max_{p} \bigg\{ R_{n,K'-K_{g}} \big(\lfloor P_{max,t}- P_{\max,g}\rfloor \big)\\+ R_{g,i}({P_{w}}^{*},{P_{s}}^{*})+ R_{g,j}({P_{s}}^{*}),\\ \{i,j\}\in K_{g}, ( P_{max,t}-P_{\max,g})\geq 0 \bigg\},
\end{multline}
where $R_{n,K'-K_{g}} \big(\lfloor P_{max,t}- P_{\max,g}\rfloor \big) $ is the maximum sum rate of the network for $ (G'-1) $ groups by allocating the remaining power $( P_{max,t}- P_{\max,g}) $.  Note that, the floor operation $ \lfloor . \rfloor  $  is used to obtain a round power value due to the fact that      the remaining power $ ( P_{max,t}- P_{\max,g}) $ might not equal to any of the power levels, and therefore,  the nearest lower level can be considered resulting in the fact that $R_{n,K'-K_{g}} \big(\lfloor P_{max,t}- P_{\max,g}\rfloor \big) $ is one of the solutions in the previous stage. It is worth mentioning  that the procedure explained above is repeated for all the levels of the power constraints until the original power constraint $ P_{max} $ in \eqref{op6} is reached, and  that the sum rate of the network is determined according to equation \eqref{op20}.
\begin{figure*}[b]
\hrule
\begin{multline}
R_{n,G'}(P_{max,t}) =
\begin {cases}
-\infty, ~~~~~~~~~~~~~ ~~~~~~~~~~~~~P_{\max,g} > P_{max,t}, g \in G ,1 \leq t \leq T, 1\leq G' \leq G \\
R_{g,i}({P_{w}}^{*},{P_{s}}^{*})+R_{g,j}({P_{s}}^{*}),~~~~~~~ P_{max,t}- P_{\max,g}=0, \{i,j\}\in K_{g} ,1 \leq t \leq T, 1\leq G' \leq G \\
\max\limits_{p} \bigg\{ R_{n,K'-K_{g}} \big(\lfloor P_{max,t}- P_{\max,g}\rfloor \big)+ R_{g,i}({P_{w}}^{*},{P_{s}}^{*})+R_{g,j}({P_{s}}^{*})\bigg\}, P_{max,t}- P_{\max,g}>0, 1 \leq t \leq T, \\~~~~~~~~~~~~~~~~~~~~~~~ ~~~~~~~~~~~~~ ~~~~~~~~~~~~~ 1\leq K' \leq K, 1\leq G' \leq G. \label{op20}
\end{cases}
\end{multline} 
\end{figure*}

Once the procedure above is applied for $ G $ groups, i.e., $ G'=G $ and $ K'=K $, the maximum sum rates of the groups are collected and stored in a set denoted by $ \mathcal{R}_{n,G} \in \mathbb{R}^{(G\times T)} $ in which the sum rate of $ K, k=\{1, \dots, K\} $, users arranged in $ G, g=\{1, \dots, G\} $, groups for each power constraint $ P_{max,t}, t=\{1, \dots, T\} $, e.g., $ R_{n,K}(P_{max,t}) $, can be found. Note that, the maximum sum rates of the $ G $  groups containing $ K $ users are determined after specifying target data rates within the range $ R_{\min,k} \leq R_{g,k}(P) \leq R_{\max,k}, $ for the weak and strong users, $ \{i,j\}\in K_{g} $, belonging to each group $ g $, which can be reached  according to a certain power allocation strategy under each power constraint.  Therefore, a set denoted by  $ \mathcal{T}_{n,K} \in \mathbb{R}^{(K\times T)} $ is assumed to host all the target data rates for the $ K $ users in the network. Moreover, a set denoted by  $ \mathcal{P}_{n,K} \in \mathbb{R}^{(K\times T)} $ is assumed to host the potential power values  for the $ K $ users in the network. Note that, the results in the set of the maximum sum rates  $ \mathcal{R}_{n,G} \in \mathbb{R}^{(G\times T)} $  are  associated with the corresponding  sets $  \mathcal{T}_{n,K} $ and $  \mathcal{P}_{n,K} $  for user-demands and power allocation, respectively. Given that, each column in the set $  \mathcal{T}_{n,K} $ is $ \left[ \mathcal{T}_{n,K}\right]_{(:,t)} $ , as well as a column in the set $  \mathcal{P}_{n,K} $ is expressed as $ \left[ \mathcal{P}_{n,K}\right]_{(:,t)} $. 

Finally after arranging the potential solutions for reaching the maximum possible sum rate in the network in three different sets  considering the group formation for the NOMA application, an optimization problem is formulated to select  the most effective solution from these sets  as follows  

\begin{subequations}
\label{op9}
\begin{align} 
\max_{\bm{G,t}} \quad & \sum_{g\in G} R_{g,K_{g}}(P_{w},P_{s})\\
\textrm{s.t.}  \quad & \sum_{k\in K} \left[ \mathcal{P}_{n,K} \right]_{(k,t)} \leq P_{max} ~~~~~~~~~\forall t\in T \label{op9_2} \\
\quad  &  R_{\min,n}\leq \sum_{k\in K} \left[ \mathcal{T}_{n,K} \right]_{(k,t)} \leq R_{\max,n}, ~~~\forall t\in T \label{op9_1} \\
\quad  & K_{g}\in K ,t \in T, g \in G, 
\end{align} 
\end{subequations} 
where $ R_{\min,n} $ and $ R_{\max,n} $ are the minimum and maximum sum rates required in the network, respectively. Interesting, this range of data rate is defined in the network to guarantee that dividing the power constraint into different levels as in \eqref{powercont} does not affect the overall energy efficiency of the network, while the power constraint is  slightly relaxed to the maximum permissible power $ P_{max} $ to consume in the network. 

The following procedure is applied to solve the optimization problem in \eqref{op9}: \begin{itemize}
\item All the maximum sum rate  solutions in the set $ \mathcal{R}_{n,G} \in \mathbb{R}^{(G\times T)} $ are considered to solve the optimization problem in \eqref{op9}, except the values that violate the constraints in  \eqref{op9_1} and \eqref{op9_2}, which are set to $ -\infty $. Similarly, the corresponding values of target data rates and power allocation in the sets $ \mathcal{T}_{n,K} $ and $ \mathcal{P}_{n,K} $ are set to 0, if and only if they do not comply with the constraints of the optimization problem in \eqref{op9}. 
\item To search for the maximum sum rate, if $ R_{n,G}(P_{max,t})=R_{n,G'}(P_{max,t}) $, while $ G'< G $, then our algorithm chooses $ R_{n,G}(P_{max,t}) $ in order to serve more groups, i.e., more users, under the same power constraint, which in turn ensures higher  spectral efficiency. On the other hand, if $ R_{n,G}(P_{max,t})=R_{n,G}(P_{max,t'}) $, while $ P_{max,t}< P_{max,t'} $, then, the proposed algorithm chooses   $ R_{n,G}(P_{max,t})$ due to the fact that the same number of users can be served at lower power consumption, which highly maximizes the energy efficiency of the network while considering the trade-offs with the spectral efficiency. In the context, the optimization problem to choose the optimum values of $ {G}^{*} $ and ${t}^{*} $  can be expressed as $ \{{G}^{*}, {t}^{*}\}=\arg\max\limits_{G,t} \sum_{g\in G} R_{g,K_{g}}(P_{w},P_{s}) $, giving higher priority for the total number of groups over the power consumption. Note that, imposing the power constraint in \eqref{op9_2} guarantees that serving the users at high power is still under the allowed power consumption in the network.  

\item The maximum sum rate $R_{n,{G}^{*}}(P_{max,{t}^{*}}) $ is achieved according to the target data rates of the $ K $ users in $  \left[ \mathcal{T}_{n,K}\right]_{(:,{t}^{*})}  $ using the power allocation strategy in $  \left[ \mathcal{P}_{n,K}\right]_{(:,{t}^{*})}  $ under the power constraint $ P_{max,{t}^{*}} $. In other words, the weak and strong users, $ \{i,j\}\in K_{g} $, belonging to each group $ g $ receive their optimum power values $ {P_{w}}^{*} $ and $ {P_{s}}^{*} $, respectively, based on the power allocation $  \left[ \mathcal{P}_{n,K}\right]_{(:,{t}^{*})}  $, which already guarantees that  the other  $(K-K_{g}) $ users belonging to other groups, $ g' \neq g $, receive  enough power to fulfill their demands determined based on $  \left[ \mathcal{T}_{n,K}\right]_{(:,{t}^{*})}  $.    
\end{itemize} 
To conclude, the original optimization problem in \eqref{op0} can be solved  by applying the proposed dynamic power allocation algorithm after the formation of multiple groups considering the principles of NOMA. In particular, the solution of the optimization problem in \eqref{op9} is  global in the network and close to the optimum solution if the above steps are applied successfully due to the fact that  the  sets, $ \mathcal{R}_{n,G} $, $ \mathcal{T}_{n,K} $ and $ \mathcal{P}_{n,K} $ are generated in accordance to the  original constraints of the main optimization problem.
\begin{table}[t]
\centering
\caption{Simulation Parameters}
\label{tabla2}
\begin{tabular}{|c|c|}
\hline
Parameter	& Value \\\hline
Transmitter Bandwidth	& 1.5 GHz \\\hline
Spectral response range  & 950-1700 nm \\\hline
Laser wavelength  & 1550 nm \\\hline
Laser beam waist & $ 8 \mu $m \\\hline
Physical area of the photodiode	&15 $\text{mm}^2$ \\\hline
Receiver FOV	& 60 deg \\\hline
Detector responsivity 	& 0.9 A/W \\\hline
Gain of optical filter & 	1.0 \\\hline
Laser noise	& $-155~ dB/H$z \\\hline
\end{tabular}
\end{table}
\section{Simulation results}
\label{sec:res}
An indoor environment with 8m$ \times $ 8m$  \times $ 3m dimensions  is considered to evaluate the performance of the dynamic application for NOMA in OWC networks. On the ceiling, $L= 4 \times 4 $ optical APs are deployed to ensure coverage for $ K=20 $ users classified as $ K_{w}=10 $  and $ K_{s}=10 $ weak and strong users, respectively. According to \cite{9803253}, the angle for full width at half maximum intensity points $ \Theta_{F} $, which affects the divergence angle $ \Theta_{D} $ and beam waist $ W_0 $,  is a vital parameter  in VCSEL that must be considered to calculate the transmit optical power that is safe for the human eye, where $ \Theta_{D}= \Theta_{F}/ \sqrt{2 \ln (2) } $, and  $ W_0=\frac{\lambda}{\pi \Theta_{D}}  $. In this context, $ 60 $ mW is set as the maximum transmit optical power per beam  for  $ \Theta_{F} =4^{\circ} $. All the other parameters are in Table~\ref{tabla2}. For comparison, two baseline schemes are proposed as counterpart schemes, which work in a similar fashion to the majority of work in the NOMA literature. 
In Baseline 1, the BIA outer precoder is used to manage  interference among multiple groups formed using the  proposed dynamic user-grouping algorithm, while a fixed power  $ P_{\max,g}=\frac{P_{\max}}{G} $ is allocated to each group in order to maximize the sum rate of each group independently through solving the optimization problem in \eqref{op8}, where $ P_{\max,t} $ is replaced by $ P_{\max,g} $.
In Baseline 2, the proposed dynamic application is used for NOMA, and the bandwidth is divided by the number of groups $ G $ to allocate an exclusive resource block for each group to avoid inter-group interference rather than determining the precoding matrices of the groups, $ \mathbf{B}= [\mathbf{B}^{[1]}, \dots, \mathbf{B}^{[g]}, \dots, \mathbf{B}^{[G]}] \in \mathbb{R}_+^{L\times G} $, following the methodology of BIA.         
\begin{figure}[t]
\begin{center}\hspace*{0cm}
\includegraphics[width=0.99\linewidth]{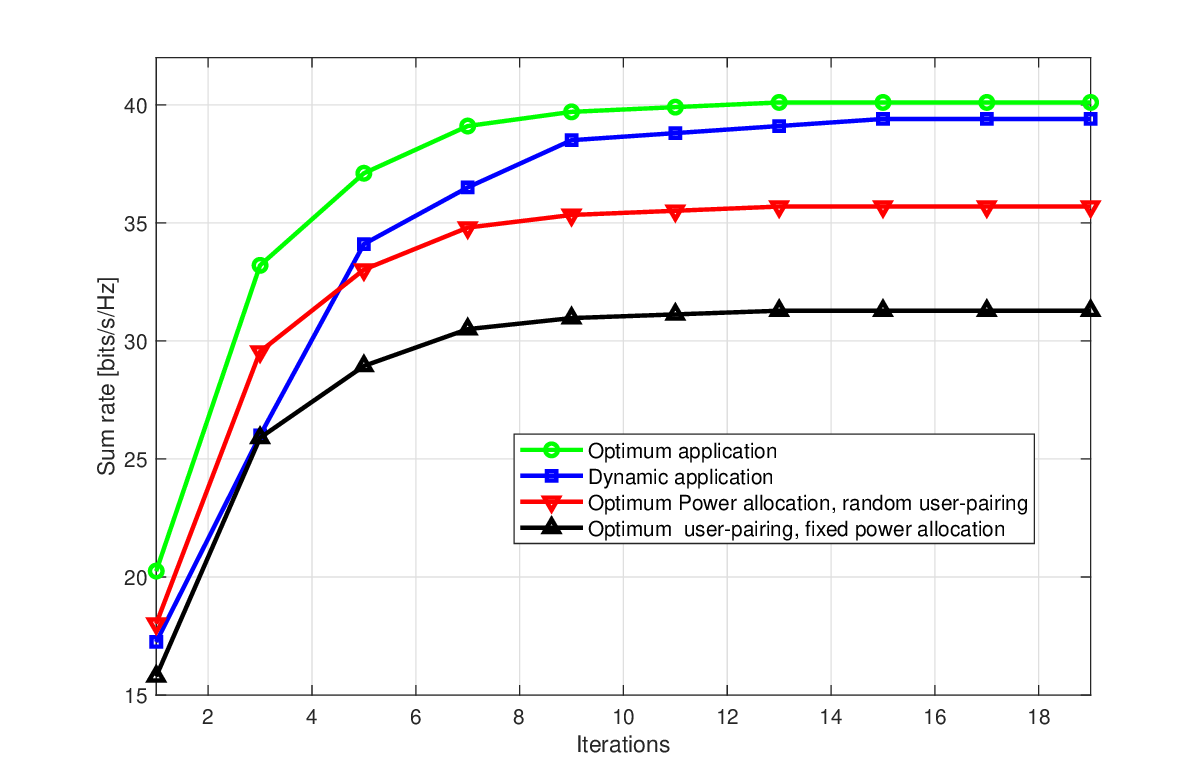}
\end{center}
\vspace{-5mm}
\caption{Sum rate versus a set of iterations for various optimization problems.}\label{fig1}
\vspace{-5mm}
\end{figure}

Fig. \ref{fig1} shows the convergence of the proposed algorithms to the optimal solution against a set of iterations. It is shown that the dynamic application of NOMA provides significant solutions close to the optimal one, which is  obtained by solving the two subproblems via exhaustive search in an alternating fashion.
The proposed  dynamic approach has   low complexity that equals to $  \mathcal{O} (T I K_{g} G)$, where $ I $ is the total number of iterations needed to maximize the sum rate of the $ K_g $ belonging  to a certain group $ g $, plus $ \mathcal{O}(T) $ of solving the optimization problem in \eqref{op9}. The figure also shows that the optimization of power allocation with random  user-paring penalizes the sum rate of the network compared to the optimal and sub-optimal solutions. Furthermore, the formation of optimum groups cannot exploit the potential of NOMA if fixed power allocation is considered. Note that, in the context of our work,  power allocation is crucial for NOMA due to the fact that the optimization problem in \eqref{op0} is formulated to maximize the total sum rate of the network and fulfill the traffic demands of the users in each group. From now on, the proposed dynamic application of NOMA is considered for the rest of the results due to its optimality at low complexity.  
\begin{figure}[t]
\begin{center}\hspace*{0cm}
\includegraphics[width=1\linewidth]{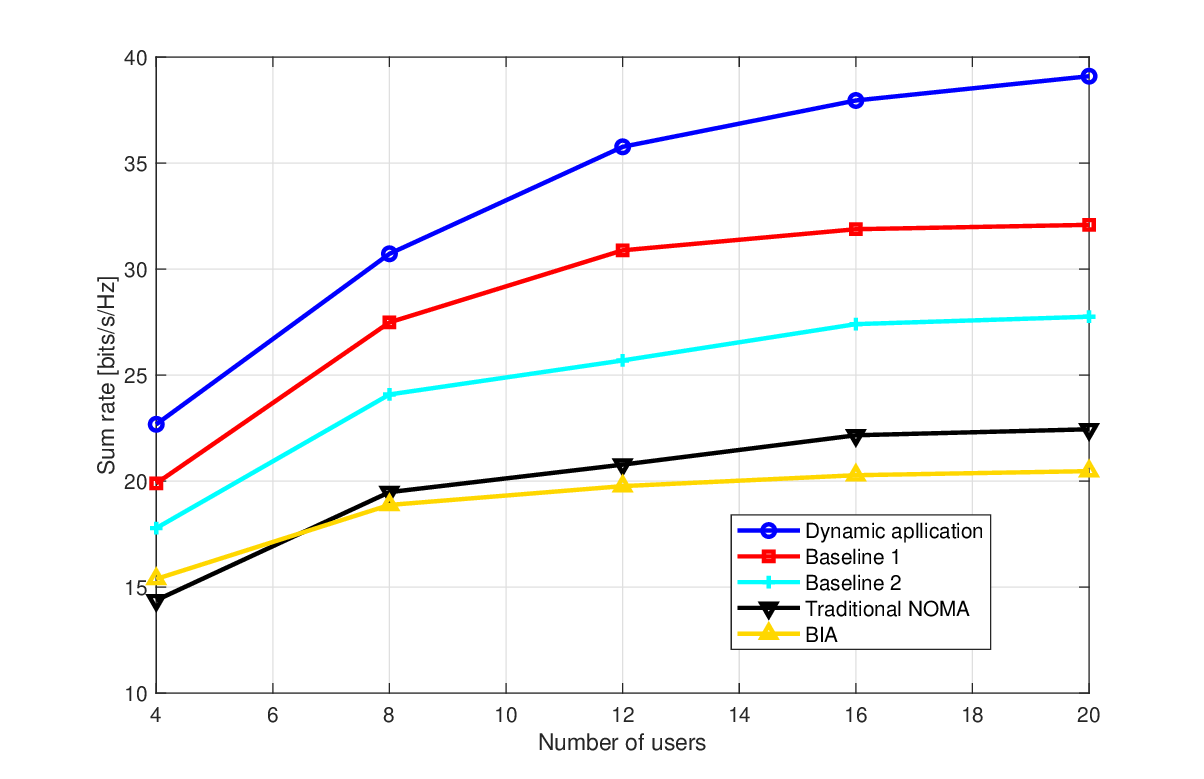}
\end{center}
\vspace{-5mm}
\caption{Sum rate versus different numbers of users for the proposed NOMA-based dynamic algorithm.}\label{fig2}
\vspace{-5mm}
\end{figure}

Fig. \ref{fig2} shows the sum rate of the network versus different numbers of users using the proposed dynamic application of NOMA compared to other benchmark schemes. It can be seen that the sum rate of the network increases with the number of users regardless of the scheme considered  as the users differ their channel gains according to their locations from the optical APs. However, the sum rate increases slightly with any number of users  beyond 15 due to the lack of the resources. The dynamic application of NOMA is superior compared to Baselines 1 and 2 due 
to the use of the  outer precoder to  avoid inter-group interference, rather than allocating orthogonal frequency or time slot to each group as in  Baseline 2, and the use of the dynamic power allocation, which  iterates over the groups to determine the optimum power values allocated to all the users in the network, compared to maximizing the sum rate of each group independently as in Baseline 1. It is worth mentioning that the conventional NOMA scheme provides low sum rates since the formation of multiple groups is obtained based on the channel gain difference between the weak and strong users, while  fixed power is allocated to each user without considering the demands of the users and the maximization of the sum rate in  the network. Moreover, BIA achieves  the lowest sum rates compared to NOMA schemes as its performance is subject to a large transmission block and high noise in high density OWC networks \cite{5734870,8636954}.

Blockage of direct optical links is a severe issue in OWC. Hence, the performance of the proposed scheme is evaluated in Fig. 4 against blockage probability. It can be seen that the sum rate of the network decreases as the blockage probability increases from 0.1 to 0.6 as  some of the weak users might be blocked from receiving service from any optical APs, and some of the  strong users might be classified as weak users. However, the proposed  dynamic application of NOMA  achieves high performance  compared to the benchmark schemes  due to the fact that the power allocated to each user is determined according to its classification and its requirements of resources sent through RF uplink transmission to guarantee high quality of service. Similarly, in Baselines 1 and 2, the power is determined according to  the demands of the users. However, the power devoted to each  group is limited in Baseline 1, and each group is served over an orthogonal transmission block in Baseline 2. It is worth pointing out that at high blockage probabilities beyond 0.6, uncovered users can experience handover to the RF network in accordance to  optimization problems with load balancing objective functions \cite{9064520}. However, it is not in the scope of the paper.

\begin{figure}[t]
\begin{center}\hspace*{0cm}
\includegraphics[width=1\linewidth]{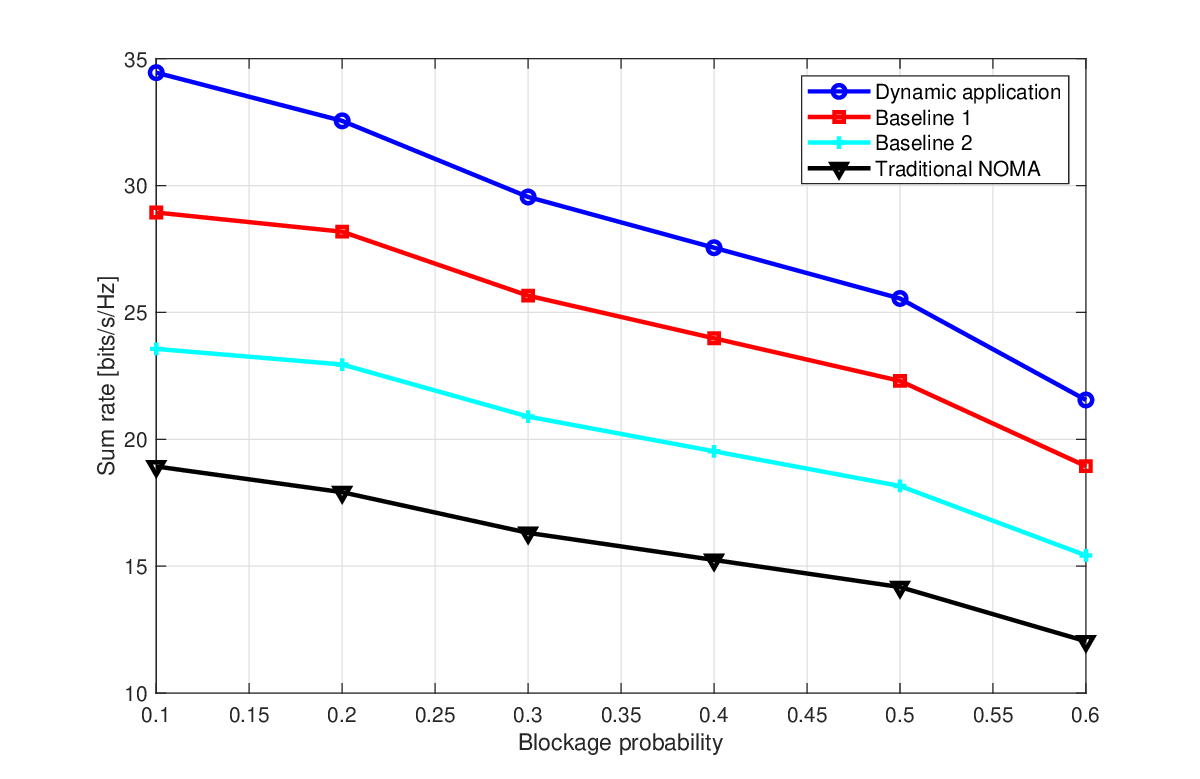}
\end{center}
\vspace{-5mm}
\caption{Sum rate versus blockage probability for different NOMA schemes.}\label{fig3}
\vspace{-5mm}
\end{figure}

\begin{figure}[t]
\begin{center}\hspace*{0cm}
\includegraphics[width=1\linewidth]{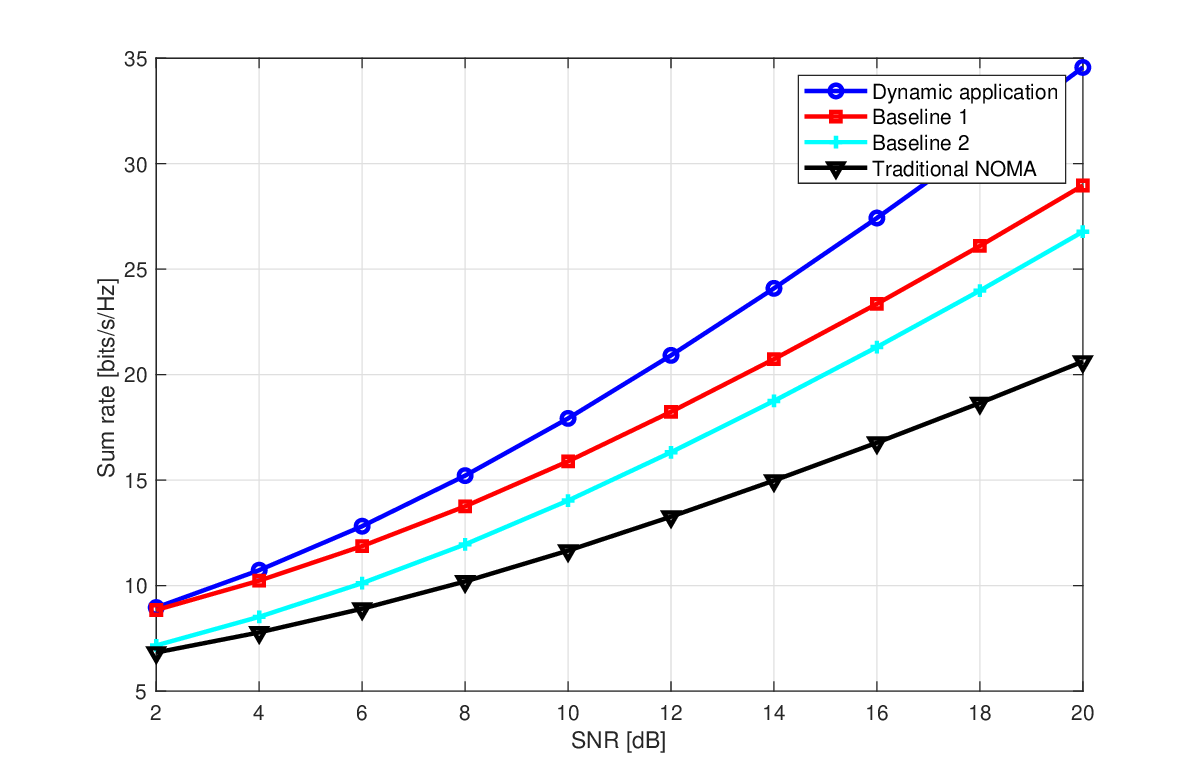}
\end{center}
\vspace{-5mm}
\caption{Sum rate versus SNR for different NOMA-based schemes.}\label{fig4}
\vspace{-5mm}
\end{figure}

Fig. \ref{fig4} shows the performance of the proposed dynamic application of NOMA against SNR. It can be seen that the proposed dynamic algorithm provides high sum rates at different SNR values due to the use of power in a more effective way, satisfying the demands of the users. In other words, if a certain user experiences low SNR due to its location, our algorithm can still satisfy its requirements if the power budget is not fully consumed. Baseline  1 shows better performance compared to Baseline 2 and the conventional NOMA  as the SNR increases. It is because of the implementation of the outer precoder for inter-group interference  and  optimizing the sum rate of the users belonging to each group. However, Baseline 1 uses  fixed power allocation strategy among the groups formed in the network, which means that if the users of a certain group demand more power to experience high quality of service, Baseline 1 might  fail to balance the load in the network causing power loss.  

\begin{figure}[t]
\begin{center}\hspace*{0cm}
\includegraphics[width=1\linewidth]{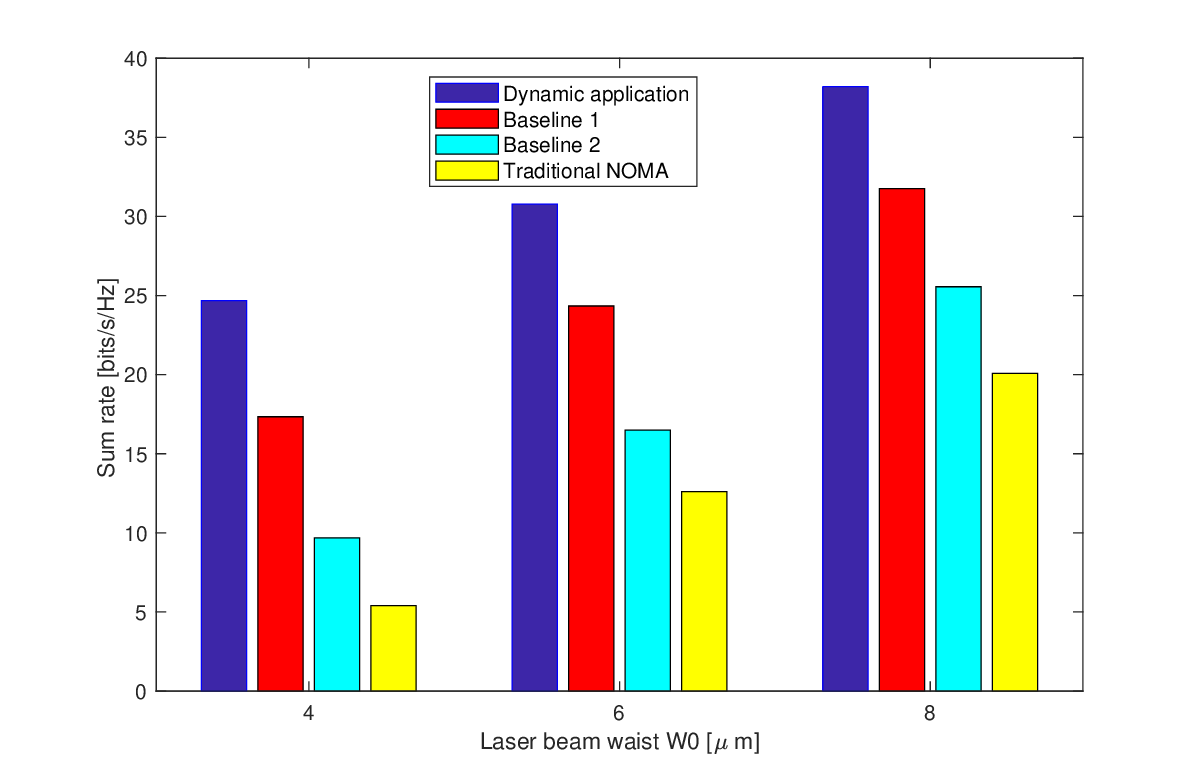}
\end{center}
\vspace{-5mm}
\caption{Sum rate versus $ W_{0} $ for different NOMA schemes.}\label{fig5}
\vspace{-5mm}
\end{figure}

\begin{figure}[t]
\begin{center}\hspace*{0cm}
\includegraphics[width=1\linewidth]{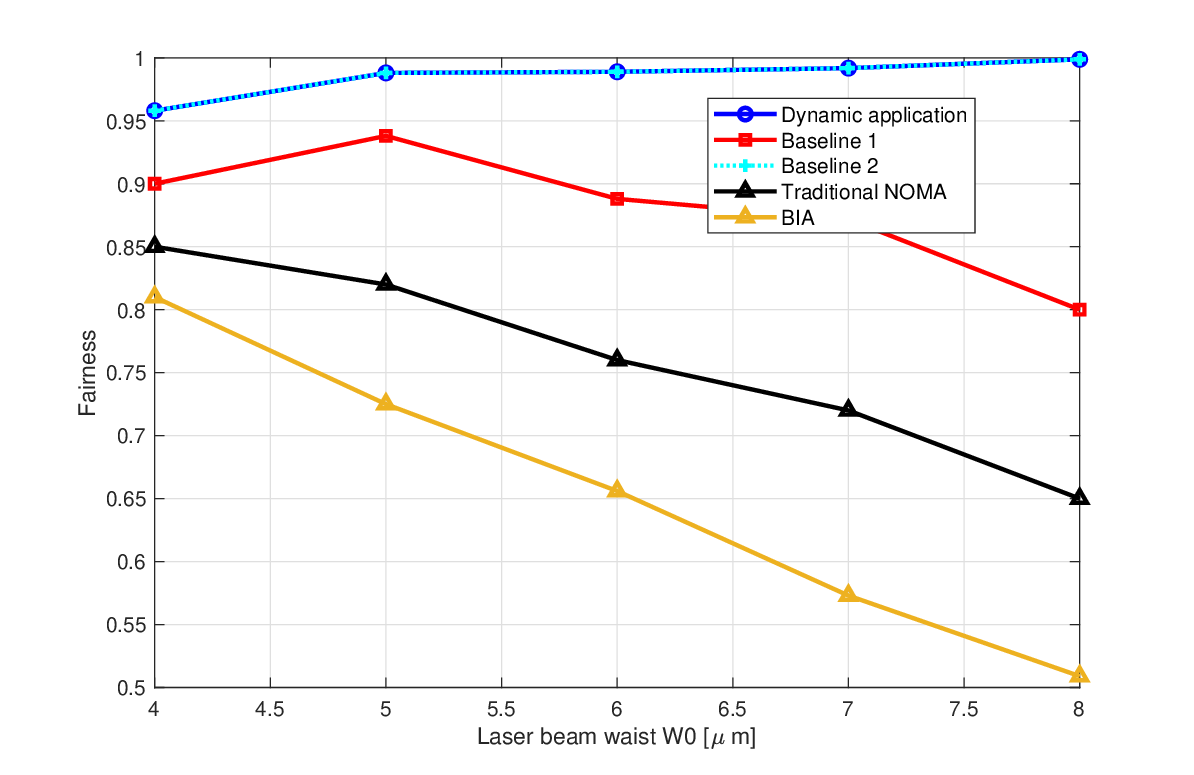}
\end{center}
\vspace{-5mm}
\caption{Fairness of the proposed dynamic algorithm versus $ W_{0} $.}\label{fig6}
\vspace{-5mm}
\end{figure}

In Fig. \ref{fig5}, the sum rate of the network is depicted against  $ W_0 $  ranging from $ 4-8 $ $ \mu $m, which determines the AP coverage area and the transmit power. It is worth mentioning  that any values beyond $ 8 $ $ \mu $m, the optical power per transmitter must be recalculated accordingly. It can be seen that the sum rate of the network increases with the beam waist due to the fact that as $ W_0 $ increases, the illuminated area of each transmitter becomes more confined focusing more power towards the users located at the center of the room, i.e., the strong users, while the weak users located at the edges of the optical cells receive low power. However, the proposed dynamic application of NOMA considers the conditions of each user and allocates power that satisfies its demands. The figure also shows the superiority of the proposed dynamic algorithm compared to the benchmark schemes. On the other hand, Fig. \ref{fig6} shows one of the important metrics in  NOMA against the beam waist, which is fairness using  Jain's fairness index as NOMA considers the channel gain of the user in power allocation. Interestingly, the fairness of the proposed scheme increases with the increase of the beam waist. It is because of, at 4 $ \mu $m,  the channel gains of the strong users are less compared to the case of 8 $ \mu $m. Note that, in the formulated optimization problem, the maximum power allocated to each strong user must be $ P_{s} \leq P^{T}_{s} $, which might not be enough for the strong user at 4 $ \mu $m beam waist. In contrast, the weak users  receive their demands in all the scenarios considered, i.e., $ P_{w} > P^{T}_{s} $. It is worth pointing out that,  Baseline 2 achieves  high fairness similar to the proposed dynamic application due to the implementation of the dynamic power allocation algorithm. While, in Baseline 1, the fairness increases with the beam waist, and it starts  to decrease with any value beyond $ 5 $ $ \mu $m. It can be interpreted that  when the beam waist increases, the weak user belonging to each group demands more power, and the power budget of each group is limited to $ P_{\max,g} $. Conventional NOMA and BIA result in poorer fairness compared to the proposed scheme, as NOMA ignores user demands and BIA allocates power equally regardless of channel gains.

\begin{figure}[t]
\begin{center}\hspace*{0cm}
\includegraphics[width=1\linewidth]{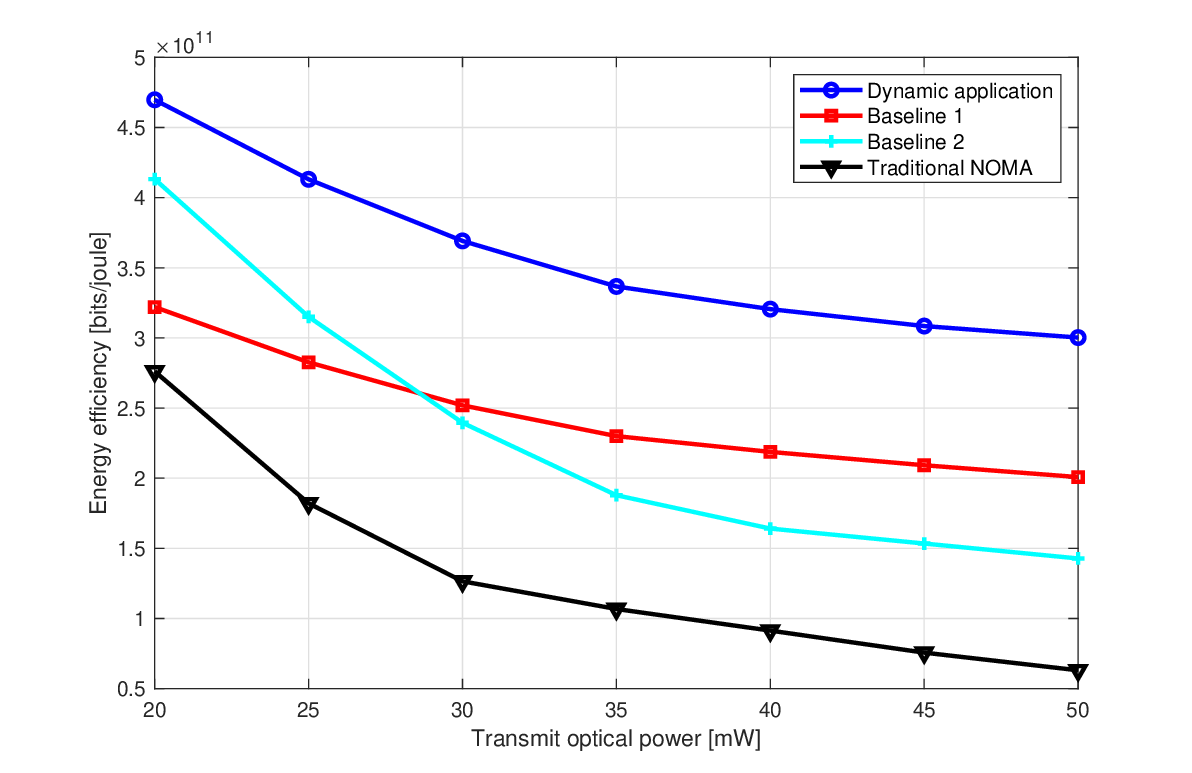}
\end{center}
\vspace{-5mm}
\caption{Energy efficiency of the proposed NOMA scheme compared to the counterpart schemes for a range of transmit optical power per transmitter.}\label{fig7}
\vspace{-5mm}
\end{figure}

In Fig. \ref{fig7}, the energy efficiency of the network is shown against the transmit power per beam to further examine the proposed dynamic use of NOMA in OWC networks. It can be seen that the energy efficiency decreases as the power increases due to power consumption. However, the dynamic application of NOMA is more efficient compared to Baselines 1 and 2 as the algorithm iterates over the groups formed to maximize the overall sum rate of the network. Interestingly, Baseline 2 overcomes Baseline 1 at low transmit power due to the implementation of the dynamic power allocation, while Baseline 1 achieves lower energy efficiency compared to the  proposed scheme due to the fact that each group receives fixed power, and users belonging to different groups might differ in their demands, and therefore, some users might not use the full power allocated to their corresponding group, and other users need more power than $ P_{\max,g} $. The conventional NOMA scheme  achieves low energy efficiency compared to the rest of the schemes  as power allocation is performed without considering the needs of the users. 
\vspace{-5mm}

\section{Conclusions}
\label{sec:con}
In this paper, a dynamic application for NOMA is proposed in a laser-based OWC network to provide high quality of service for multiple users arranged into groups. We first define the system model consisting of multiple optical APs, where each AP is an array of VCSELs, derive the optical channel to determine user connectivity, and introduce  optical constraints to ensure adherence to eye safety regulations. Then, a BIA outer precoder is designed for AP coordination and inter-group interference management, enabling the application of NOMA in each group. Finally, an optimization problem is formulated to  maximize the sum rate of the network though jointly optimizing weak-strong user combinations and power allocation, while fulfilling user traffic demands. The main optimization problem is  a max-min fractional program, which is  difficult to
solve in practice. Therefore, two dynamic algorithms are designed to solve the user grouping problem and power allocation separately with the guarantee of convergence to the optimal solution.  
The results show the superiority of the proposed dynamic application of NOMA in terms of data rate, energy efficiency, and fairness compared to counterpart schemes introduced according to the literature of NOMA. For future work, 
BIA outer precoders have limitations in terms of transmission block size and inherent noise, and therefore, new precoder structures can be determined by formulating rate-maximization and OWC-RF load balancing problems under the  constraints of BIA and NOMA. Additionally, machine learning algorithms can be implemented to provide solutions to these complex problems with real-time adaptability.
\appendix
By finding a root of equation $ \mathcal{L}(\xi_{g})=0  $ using  a Dinkelbach-based algorithm \cite{2009convex} under the constraints of NOMA in \eqref{op8}, the sum rate of group $ g $ can be maximized.  The Lagrangian function of \eqref{di} for  weak user $ i\in K_{g} $  belonging to group $ g $ under the constraints in \eqref{op8} can be expressed as   
\begin{multline}
\label{lag}
\mathcal{L}_{i}= R_{g,i}(P_{w},{P^{T}_{s}})- \xi_{g} (P_{w}+{P^{T}_{s}})+ \alpha_{g} (P_{\max,g}-(P_{w}+{P^{T}_{s}}))\\+\mu_{g,i}(P_{w}-{P^{T}_{s}})+\lambda_{\max,i}(R_{\max,k}-R_{g,i}(P_{w},{P^{T}_{s}}))\\+ \nu_{\min,i} (R_{g,i}(P_{w},{P^{T}_{s}})-R_{\min,k}),
\end{multline}  
where $ \alpha_{g} $, $ \mu_{g,i} $,  $ \lambda_{\max,i} $ and $ \nu_{\min,i} $  are Lagrangian multipliers defined to determine the  power allocated to the weak user according to the constraints in \eqref{op8_1}, 
\eqref{op8_2} and \eqref{op8_3}, respectively.
Using the Karush-Kuhn-Tucker (KKT) conditions \cite{9521837,2009convex}, the partial derivative $ \frac{ \partial \mathcal{L}_{i}}{\partial P_{w} } =0$,i.e., 
\begin{equation}
\frac{ \partial \mathcal{L}_{i}}{\partial P_{w} }= (\nu_{\min,i}-\lambda_{\max,i} +1)\frac{ R_{g,i}(P_{w}, {P^{T}_{s}})}{\partial P_{w} }- \xi_{g} + \alpha_{g}-\mu_{g,i}=0. 
\end{equation}
At given values  for $  \xi_{g} $, $ \alpha_{g} $, $ \mu_{g,i} $,  $ \lambda_{\max,i} $ and $ \nu_{\min,i} $, the optimum power ${ P_{w}}^{*} $ for the weak user, $ i\in K_{g} $, can be found,  and equation \eqref{lag} can be updated considering that optimum power. Subsequently, the dual problem can be solved using the gradient projection method, and  the Lagrangian
multipliers related to power allocation  and user demands are updated as follows 
\begin{equation}
\label{var1}
\mu_{g,i} (\tau) = \left[\mu_{g,i} (\tau-1)-\epsilon_{1}(\tau-1) \left(P_{w}(\tau-1)-P^{T}_{s} \right) \right]^{+},
\end{equation} 
\begin{multline}
\label{var2}
\alpha_{g} (\tau)= \bigg[\alpha_{g} (\tau-1)-\epsilon_{2}(\tau-1) \Big(P_{\max,g} - \Big(P_{w}(\tau-1) +P^{T}_{s}\Big)\Big) \bigg]^{+},
\end{multline}
\begin{multline}
\label{var4}
\lambda_{\max,i} (\tau)= \bigg[\lambda_{\max,i} (\tau-1)-\epsilon_{3}(\tau-1) \\ \bigg(R_{\max,k}  - R_{g,i}(P_{w},{P^{T}_{s}}) (\tau-1) \bigg) \bigg]^{+},
\end{multline} 
 
\begin{multline}
\label{var5}
\nu_{\min,i} (\tau)= \bigg[\nu_{\min,i} (\tau-1)-\epsilon_{4}(\tau-1) \\ \bigg( R_{g,i}(P_{w},{P^{T}_{s}})(\tau-1)- R_{\min,k} \bigg) \bigg]^{+},
\end{multline} 
where $ \tau$ denotes the iteration of the gradient algorithm, and  $ [.]^{+} $ is a projection on the positive orthant taking into consideration that $ \mu_{g,i} , \alpha_{g}, \lambda_{\max,i}, \nu_{\min,i} \geq 0 $. Moreover, $ \epsilon_{n}(\tau-1) $, $ n=\{1,2,3,4\} $,  
is a sufficient small step size  at a certain iteration.  The multipliers  $ \mu_{g,i} (\tau) $ and $ \alpha_{g} $ are updated iteratively until the power allocated to weak user $ i\in K_g $  is determined in compliance with the constraints \eqref{op8_1} and \eqref{op8_2}, and the
multipliers  $ \lambda_{\max,i} $ and $ \nu_{\min,i} $ are  updated to  guarantee that  the data rate of the weak user achieved by allocating $ { P_{w}}^{*}  $  at $ {\mu_{g,i}}^{*} $ and $ {\alpha_{g}}^{*} $ is enough to satisfy its requirements. The overall algorithm iterates over power allocation until a closed-form solution is reached. It is worth pointing out that the same methodology can be easily followed to determine the power allocated to strong user $ j\in K_{g} $ belonging to group $ g $ considering the constraint $  P_{s} \leq P^{T}_{s}$ rather than the constraint   
$ P_{w} > P^{T}_{s}$.

\bibliographystyle{IEEEtran}
\bibliography{mybib}
\end{document}